\documentclass[12pt]{article}
\usepackage{epsfig,amsfonts,amssymb}
\usepackage{hyperref}
\usepackage{cite}
\input epsf.sty
\usepackage{cite}

\def\ZZZ{{\hbox{ Z\kern-1.6mm Z}}}
\def\RRR{{\hbox{ R\kern-2.4mm R}}}
\def\CCC{{\hbox{ C\kern-2.0mm C}}}
\def\zzz{{\hbox{z\kern-1mm z}}}

\newcommand{\qeq}{{\hbox{=\kern-2.3mm ? \kern.5mm }}}
\renewcommand{\qeq}{=}

\newcommand{\ve}{\varepsilon}

\newcommand{\II}{{\cal I}}

\newcommand{\FF}{{\cal F}}

\newcommand{\HH}{{\cal H}}
\newcommand{\MM}{{\cal M}}

\newcommand{\OO}{{\cal O}}

\newcommand{\PP}{{\cal V}}

\newcommand{\LL}{{\cal L}}

\newcommand{\wt}{\widetilde}
\newcommand{\wh}{\widehat}

\newcommand{\be}{\begin{equation}}
\newcommand{\ee}{\end{equation}}
\newcommand{\ben}{\begin{eqnarray}\displaystyle}
\newcommand{\een}{\end{eqnarray}}

\newcommand{\refb}[1]{(\ref{#1})}
\newcommand{\p}{\partial}
\newcommand{\sectiono}[1]{\section{#1}\setcounter{equation}{0}}

\def\one{{\hbox{ 1\kern-.8mm l}}}
\def\zero{{\hbox{ 0\kern-1.5mm 0}}}

\newcommand{\bi}{{\bf i}}

\newcommand{\bea}[1]{\begin{eqnarray}\label{#1} }
\newcommand{\eea}{\end{eqnarray}}

\newcommand{\eqref}{\refb}

\usepackage{bm}
\usepackage[table]{xcolor}

\def\figa{
\def\JPicScale{0.8}
\ifx\JPicScale\undefined\def\JPicScale{1}\fi
\unitlength \JPicScale mm
\begin{picture}(135,42.5)(0,0)
\linethickness{0.3mm}
\put(10,40){\line(1,0){10}}
\linethickness{0.3mm}
\put(40,39.94){\line(0,1){0.13}}
\multiput(39.97,40.19)(0.03,-0.13){1}{\line(0,-1){0.13}}
\multiput(39.92,40.32)(0.05,-0.13){1}{\line(0,-1){0.13}}
\multiput(39.84,40.44)(0.08,-0.13){1}{\line(0,-1){0.13}}
\multiput(39.74,40.57)(0.1,-0.12){1}{\line(0,-1){0.12}}
\multiput(39.61,40.69)(0.13,-0.12){1}{\line(1,0){0.13}}
\multiput(39.46,40.81)(0.15,-0.12){1}{\line(1,0){0.15}}
\multiput(39.29,40.93)(0.18,-0.12){1}{\line(1,0){0.18}}
\multiput(39.09,41.04)(0.2,-0.12){1}{\line(1,0){0.2}}
\multiput(38.86,41.16)(0.22,-0.11){1}{\line(1,0){0.22}}
\multiput(38.62,41.27)(0.25,-0.11){1}{\line(1,0){0.25}}
\multiput(38.35,41.38)(0.27,-0.11){1}{\line(1,0){0.27}}
\multiput(38.06,41.48)(0.29,-0.1){1}{\line(1,0){0.29}}
\multiput(37.75,41.58)(0.31,-0.1){1}{\line(1,0){0.31}}
\multiput(37.42,41.68)(0.33,-0.1){1}{\line(1,0){0.33}}
\multiput(37.07,41.77)(0.35,-0.09){1}{\line(1,0){0.35}}
\multiput(36.7,41.86)(0.37,-0.09){1}{\line(1,0){0.37}}
\multiput(36.32,41.94)(0.38,-0.08){1}{\line(1,0){0.38}}
\multiput(35.92,42.02)(0.4,-0.08){1}{\line(1,0){0.4}}
\multiput(35.5,42.09)(0.42,-0.07){1}{\line(1,0){0.42}}
\multiput(35.07,42.15)(0.43,-0.07){1}{\line(1,0){0.43}}
\multiput(34.63,42.22)(0.44,-0.06){1}{\line(1,0){0.44}}
\multiput(34.18,42.27)(0.45,-0.06){1}{\line(1,0){0.45}}
\multiput(33.71,42.32)(0.47,-0.05){1}{\line(1,0){0.47}}
\multiput(33.23,42.37)(0.48,-0.04){1}{\line(1,0){0.48}}
\multiput(32.75,42.4)(0.48,-0.04){1}{\line(1,0){0.48}}
\multiput(32.26,42.44)(0.49,-0.03){1}{\line(1,0){0.49}}
\multiput(31.76,42.46)(0.5,-0.03){1}{\line(1,0){0.5}}
\multiput(31.26,42.48)(0.5,-0.02){1}{\line(1,0){0.5}}
\multiput(30.76,42.49)(0.5,-0.01){1}{\line(1,0){0.5}}
\multiput(30.25,42.5)(0.51,-0.01){1}{\line(1,0){0.51}}
\put(29.75,42.5){\line(1,0){0.51}}
\multiput(29.24,42.49)(0.51,0.01){1}{\line(1,0){0.51}}
\multiput(28.74,42.48)(0.5,0.01){1}{\line(1,0){0.5}}
\multiput(28.24,42.46)(0.5,0.02){1}{\line(1,0){0.5}}
\multiput(27.74,42.44)(0.5,0.03){1}{\line(1,0){0.5}}
\multiput(27.25,42.4)(0.49,0.03){1}{\line(1,0){0.49}}
\multiput(26.77,42.37)(0.48,0.04){1}{\line(1,0){0.48}}
\multiput(26.29,42.32)(0.48,0.04){1}{\line(1,0){0.48}}
\multiput(25.82,42.27)(0.47,0.05){1}{\line(1,0){0.47}}
\multiput(25.37,42.22)(0.45,0.06){1}{\line(1,0){0.45}}
\multiput(24.93,42.15)(0.44,0.06){1}{\line(1,0){0.44}}
\multiput(24.5,42.09)(0.43,0.07){1}{\line(1,0){0.43}}
\multiput(24.08,42.02)(0.42,0.07){1}{\line(1,0){0.42}}
\multiput(23.68,41.94)(0.4,0.08){1}{\line(1,0){0.4}}
\multiput(23.3,41.86)(0.38,0.08){1}{\line(1,0){0.38}}
\multiput(22.93,41.77)(0.37,0.09){1}{\line(1,0){0.37}}
\multiput(22.58,41.68)(0.35,0.09){1}{\line(1,0){0.35}}
\multiput(22.25,41.58)(0.33,0.1){1}{\line(1,0){0.33}}
\multiput(21.94,41.48)(0.31,0.1){1}{\line(1,0){0.31}}
\multiput(21.65,41.38)(0.29,0.1){1}{\line(1,0){0.29}}
\multiput(21.38,41.27)(0.27,0.11){1}{\line(1,0){0.27}}
\multiput(21.14,41.16)(0.25,0.11){1}{\line(1,0){0.25}}
\multiput(20.91,41.04)(0.22,0.11){1}{\line(1,0){0.22}}
\multiput(20.71,40.93)(0.2,0.12){1}{\line(1,0){0.2}}
\multiput(20.54,40.81)(0.18,0.12){1}{\line(1,0){0.18}}
\multiput(20.39,40.69)(0.15,0.12){1}{\line(1,0){0.15}}
\multiput(20.26,40.57)(0.13,0.12){1}{\line(1,0){0.13}}
\multiput(20.16,40.44)(0.1,0.12){1}{\line(0,1){0.12}}
\multiput(20.08,40.32)(0.08,0.13){1}{\line(0,1){0.13}}
\multiput(20.03,40.19)(0.05,0.13){1}{\line(0,1){0.13}}
\multiput(20,40.06)(0.03,0.13){1}{\line(0,1){0.13}}
\put(20,39.94){\line(0,1){0.13}}
\multiput(20,39.94)(0.03,-0.13){1}{\line(0,-1){0.13}}
\multiput(20.03,39.81)(0.05,-0.13){1}{\line(0,-1){0.13}}
\multiput(20.08,39.68)(0.08,-0.13){1}{\line(0,-1){0.13}}
\multiput(20.16,39.56)(0.1,-0.12){1}{\line(0,-1){0.12}}
\multiput(20.26,39.43)(0.13,-0.12){1}{\line(1,0){0.13}}
\multiput(20.39,39.31)(0.15,-0.12){1}{\line(1,0){0.15}}
\multiput(20.54,39.19)(0.18,-0.12){1}{\line(1,0){0.18}}
\multiput(20.71,39.07)(0.2,-0.12){1}{\line(1,0){0.2}}
\multiput(20.91,38.96)(0.22,-0.11){1}{\line(1,0){0.22}}
\multiput(21.14,38.84)(0.25,-0.11){1}{\line(1,0){0.25}}
\multiput(21.38,38.73)(0.27,-0.11){1}{\line(1,0){0.27}}
\multiput(21.65,38.62)(0.29,-0.1){1}{\line(1,0){0.29}}
\multiput(21.94,38.52)(0.31,-0.1){1}{\line(1,0){0.31}}
\multiput(22.25,38.42)(0.33,-0.1){1}{\line(1,0){0.33}}
\multiput(22.58,38.32)(0.35,-0.09){1}{\line(1,0){0.35}}
\multiput(22.93,38.23)(0.37,-0.09){1}{\line(1,0){0.37}}
\multiput(23.3,38.14)(0.38,-0.08){1}{\line(1,0){0.38}}
\multiput(23.68,38.06)(0.4,-0.08){1}{\line(1,0){0.4}}
\multiput(24.08,37.98)(0.42,-0.07){1}{\line(1,0){0.42}}
\multiput(24.5,37.91)(0.43,-0.07){1}{\line(1,0){0.43}}
\multiput(24.93,37.85)(0.44,-0.06){1}{\line(1,0){0.44}}
\multiput(25.37,37.78)(0.45,-0.06){1}{\line(1,0){0.45}}
\multiput(25.82,37.73)(0.47,-0.05){1}{\line(1,0){0.47}}
\multiput(26.29,37.68)(0.48,-0.04){1}{\line(1,0){0.48}}
\multiput(26.77,37.63)(0.48,-0.04){1}{\line(1,0){0.48}}
\multiput(27.25,37.6)(0.49,-0.03){1}{\line(1,0){0.49}}
\multiput(27.74,37.56)(0.5,-0.03){1}{\line(1,0){0.5}}
\multiput(28.24,37.54)(0.5,-0.02){1}{\line(1,0){0.5}}
\multiput(28.74,37.52)(0.5,-0.01){1}{\line(1,0){0.5}}
\multiput(29.24,37.51)(0.51,-0.01){1}{\line(1,0){0.51}}
\put(29.75,37.5){\line(1,0){0.51}}
\multiput(30.25,37.5)(0.51,0.01){1}{\line(1,0){0.51}}
\multiput(30.76,37.51)(0.5,0.01){1}{\line(1,0){0.5}}
\multiput(31.26,37.52)(0.5,0.02){1}{\line(1,0){0.5}}
\multiput(31.76,37.54)(0.5,0.03){1}{\line(1,0){0.5}}
\multiput(32.26,37.56)(0.49,0.03){1}{\line(1,0){0.49}}
\multiput(32.75,37.6)(0.48,0.04){1}{\line(1,0){0.48}}
\multiput(33.23,37.63)(0.48,0.04){1}{\line(1,0){0.48}}
\multiput(33.71,37.68)(0.47,0.05){1}{\line(1,0){0.47}}
\multiput(34.18,37.73)(0.45,0.06){1}{\line(1,0){0.45}}
\multiput(34.63,37.78)(0.44,0.06){1}{\line(1,0){0.44}}
\multiput(35.07,37.85)(0.43,0.07){1}{\line(1,0){0.43}}
\multiput(35.5,37.91)(0.42,0.07){1}{\line(1,0){0.42}}
\multiput(35.92,37.98)(0.4,0.08){1}{\line(1,0){0.4}}
\multiput(36.32,38.06)(0.38,0.08){1}{\line(1,0){0.38}}
\multiput(36.7,38.14)(0.37,0.09){1}{\line(1,0){0.37}}
\multiput(37.07,38.23)(0.35,0.09){1}{\line(1,0){0.35}}
\multiput(37.42,38.32)(0.33,0.1){1}{\line(1,0){0.33}}
\multiput(37.75,38.42)(0.31,0.1){1}{\line(1,0){0.31}}
\multiput(38.06,38.52)(0.29,0.1){1}{\line(1,0){0.29}}
\multiput(38.35,38.62)(0.27,0.11){1}{\line(1,0){0.27}}
\multiput(38.62,38.73)(0.25,0.11){1}{\line(1,0){0.25}}
\multiput(38.86,38.84)(0.22,0.11){1}{\line(1,0){0.22}}
\multiput(39.09,38.96)(0.2,0.12){1}{\line(1,0){0.2}}
\multiput(39.29,39.07)(0.18,0.12){1}{\line(1,0){0.18}}
\multiput(39.46,39.19)(0.15,0.12){1}{\line(1,0){0.15}}
\multiput(39.61,39.31)(0.13,0.12){1}{\line(1,0){0.13}}
\multiput(39.74,39.43)(0.1,0.12){1}{\line(0,1){0.12}}
\multiput(39.84,39.56)(0.08,0.13){1}{\line(0,1){0.13}}
\multiput(39.92,39.68)(0.05,0.13){1}{\line(0,1){0.13}}
\multiput(39.97,39.81)(0.03,0.13){1}{\line(0,1){0.13}}
\linethickness{0.3mm}
\put(-20,40){\line(1,0){15}}
\linethickness{0.3mm}
\put(40,40){\line(1,0){10}}
\linethickness{0.3mm}
\put(60,40){\line(1,0){10}}
\linethickness{0.3mm}
\put(90,40){\line(1,0){10}}
\linethickness{0.3mm}
\put(120,40){\line(1,0){10}}
\linethickness{0.3mm}
\put(120,39.94){\line(0,1){0.13}}
\multiput(119.97,40.19)(0.03,-0.13){1}{\line(0,-1){0.13}}
\multiput(119.92,40.32)(0.05,-0.13){1}{\line(0,-1){0.13}}
\multiput(119.84,40.44)(0.08,-0.13){1}{\line(0,-1){0.13}}
\multiput(119.74,40.57)(0.1,-0.12){1}{\line(0,-1){0.12}}
\multiput(119.61,40.69)(0.13,-0.12){1}{\line(1,0){0.13}}
\multiput(119.46,40.81)(0.15,-0.12){1}{\line(1,0){0.15}}
\multiput(119.29,40.93)(0.18,-0.12){1}{\line(1,0){0.18}}
\multiput(119.09,41.04)(0.2,-0.12){1}{\line(1,0){0.2}}
\multiput(118.86,41.16)(0.22,-0.11){1}{\line(1,0){0.22}}
\multiput(118.62,41.27)(0.25,-0.11){1}{\line(1,0){0.25}}
\multiput(118.35,41.38)(0.27,-0.11){1}{\line(1,0){0.27}}
\multiput(118.06,41.48)(0.29,-0.1){1}{\line(1,0){0.29}}
\multiput(117.75,41.58)(0.31,-0.1){1}{\line(1,0){0.31}}
\multiput(117.42,41.68)(0.33,-0.1){1}{\line(1,0){0.33}}
\multiput(117.07,41.77)(0.35,-0.09){1}{\line(1,0){0.35}}
\multiput(116.7,41.86)(0.37,-0.09){1}{\line(1,0){0.37}}
\multiput(116.32,41.94)(0.38,-0.08){1}{\line(1,0){0.38}}
\multiput(115.92,42.02)(0.4,-0.08){1}{\line(1,0){0.4}}
\multiput(115.5,42.09)(0.42,-0.07){1}{\line(1,0){0.42}}
\multiput(115.07,42.15)(0.43,-0.07){1}{\line(1,0){0.43}}
\multiput(114.63,42.22)(0.44,-0.06){1}{\line(1,0){0.44}}
\multiput(114.18,42.27)(0.45,-0.06){1}{\line(1,0){0.45}}
\multiput(113.71,42.32)(0.47,-0.05){1}{\line(1,0){0.47}}
\multiput(113.23,42.37)(0.48,-0.04){1}{\line(1,0){0.48}}
\multiput(112.75,42.4)(0.48,-0.04){1}{\line(1,0){0.48}}
\multiput(112.26,42.44)(0.49,-0.03){1}{\line(1,0){0.49}}
\multiput(111.76,42.46)(0.5,-0.03){1}{\line(1,0){0.5}}
\multiput(111.26,42.48)(0.5,-0.02){1}{\line(1,0){0.5}}
\multiput(110.76,42.49)(0.5,-0.01){1}{\line(1,0){0.5}}
\multiput(110.25,42.5)(0.51,-0.01){1}{\line(1,0){0.51}}
\put(109.75,42.5){\line(1,0){0.51}}
\multiput(109.24,42.49)(0.51,0.01){1}{\line(1,0){0.51}}
\multiput(108.74,42.48)(0.5,0.01){1}{\line(1,0){0.5}}
\multiput(108.24,42.46)(0.5,0.02){1}{\line(1,0){0.5}}
\multiput(107.74,42.44)(0.5,0.03){1}{\line(1,0){0.5}}
\multiput(107.25,42.4)(0.49,0.03){1}{\line(1,0){0.49}}
\multiput(106.77,42.37)(0.48,0.04){1}{\line(1,0){0.48}}
\multiput(106.29,42.32)(0.48,0.04){1}{\line(1,0){0.48}}
\multiput(105.82,42.27)(0.47,0.05){1}{\line(1,0){0.47}}
\multiput(105.37,42.22)(0.45,0.06){1}{\line(1,0){0.45}}
\multiput(104.93,42.15)(0.44,0.06){1}{\line(1,0){0.44}}
\multiput(104.5,42.09)(0.43,0.07){1}{\line(1,0){0.43}}
\multiput(104.08,42.02)(0.42,0.07){1}{\line(1,0){0.42}}
\multiput(103.68,41.94)(0.4,0.08){1}{\line(1,0){0.4}}
\multiput(103.3,41.86)(0.38,0.08){1}{\line(1,0){0.38}}
\multiput(102.93,41.77)(0.37,0.09){1}{\line(1,0){0.37}}
\multiput(102.58,41.68)(0.35,0.09){1}{\line(1,0){0.35}}
\multiput(102.25,41.58)(0.33,0.1){1}{\line(1,0){0.33}}
\multiput(101.94,41.48)(0.31,0.1){1}{\line(1,0){0.31}}
\multiput(101.65,41.38)(0.29,0.1){1}{\line(1,0){0.29}}
\multiput(101.38,41.27)(0.27,0.11){1}{\line(1,0){0.27}}
\multiput(101.14,41.16)(0.25,0.11){1}{\line(1,0){0.25}}
\multiput(100.91,41.04)(0.22,0.11){1}{\line(1,0){0.22}}
\multiput(100.71,40.93)(0.2,0.12){1}{\line(1,0){0.2}}
\multiput(100.54,40.81)(0.18,0.12){1}{\line(1,0){0.18}}
\multiput(100.39,40.69)(0.15,0.12){1}{\line(1,0){0.15}}
\multiput(100.26,40.57)(0.13,0.12){1}{\line(1,0){0.13}}
\multiput(100.16,40.44)(0.1,0.12){1}{\line(0,1){0.12}}
\multiput(100.08,40.32)(0.08,0.13){1}{\line(0,1){0.13}}
\multiput(100.03,40.19)(0.05,0.13){1}{\line(0,1){0.13}}
\multiput(100,40.06)(0.03,0.13){1}{\line(0,1){0.13}}
\put(100,39.94){\line(0,1){0.13}}
\multiput(100,39.94)(0.03,-0.13){1}{\line(0,-1){0.13}}
\multiput(100.03,39.81)(0.05,-0.13){1}{\line(0,-1){0.13}}
\multiput(100.08,39.68)(0.08,-0.13){1}{\line(0,-1){0.13}}
\multiput(100.16,39.56)(0.1,-0.12){1}{\line(0,-1){0.12}}
\multiput(100.26,39.43)(0.13,-0.12){1}{\line(1,0){0.13}}
\multiput(100.39,39.31)(0.15,-0.12){1}{\line(1,0){0.15}}
\multiput(100.54,39.19)(0.18,-0.12){1}{\line(1,0){0.18}}
\multiput(100.71,39.07)(0.2,-0.12){1}{\line(1,0){0.2}}
\multiput(100.91,38.96)(0.22,-0.11){1}{\line(1,0){0.22}}
\multiput(101.14,38.84)(0.25,-0.11){1}{\line(1,0){0.25}}
\multiput(101.38,38.73)(0.27,-0.11){1}{\line(1,0){0.27}}
\multiput(101.65,38.62)(0.29,-0.1){1}{\line(1,0){0.29}}
\multiput(101.94,38.52)(0.31,-0.1){1}{\line(1,0){0.31}}
\multiput(102.25,38.42)(0.33,-0.1){1}{\line(1,0){0.33}}
\multiput(102.58,38.32)(0.35,-0.09){1}{\line(1,0){0.35}}
\multiput(102.93,38.23)(0.37,-0.09){1}{\line(1,0){0.37}}
\multiput(103.3,38.14)(0.38,-0.08){1}{\line(1,0){0.38}}
\multiput(103.68,38.06)(0.4,-0.08){1}{\line(1,0){0.4}}
\multiput(104.08,37.98)(0.42,-0.07){1}{\line(1,0){0.42}}
\multiput(104.5,37.91)(0.43,-0.07){1}{\line(1,0){0.43}}
\multiput(104.93,37.85)(0.44,-0.06){1}{\line(1,0){0.44}}
\multiput(105.37,37.78)(0.45,-0.06){1}{\line(1,0){0.45}}
\multiput(105.82,37.73)(0.47,-0.05){1}{\line(1,0){0.47}}
\multiput(106.29,37.68)(0.48,-0.04){1}{\line(1,0){0.48}}
\multiput(106.77,37.63)(0.48,-0.04){1}{\line(1,0){0.48}}
\multiput(107.25,37.6)(0.49,-0.03){1}{\line(1,0){0.49}}
\multiput(107.74,37.56)(0.5,-0.03){1}{\line(1,0){0.5}}
\multiput(108.24,37.54)(0.5,-0.02){1}{\line(1,0){0.5}}
\multiput(108.74,37.52)(0.5,-0.01){1}{\line(1,0){0.5}}
\multiput(109.24,37.51)(0.51,-0.01){1}{\line(1,0){0.51}}
\put(109.75,37.5){\line(1,0){0.51}}
\multiput(110.25,37.5)(0.51,0.01){1}{\line(1,0){0.51}}
\multiput(110.76,37.51)(0.5,0.01){1}{\line(1,0){0.5}}
\multiput(111.26,37.52)(0.5,0.02){1}{\line(1,0){0.5}}
\multiput(111.76,37.54)(0.5,0.03){1}{\line(1,0){0.5}}
\multiput(112.26,37.56)(0.49,0.03){1}{\line(1,0){0.49}}
\multiput(112.75,37.6)(0.48,0.04){1}{\line(1,0){0.48}}
\multiput(113.23,37.63)(0.48,0.04){1}{\line(1,0){0.48}}
\multiput(113.71,37.68)(0.47,0.05){1}{\line(1,0){0.47}}
\multiput(114.18,37.73)(0.45,0.06){1}{\line(1,0){0.45}}
\multiput(114.63,37.78)(0.44,0.06){1}{\line(1,0){0.44}}
\multiput(115.07,37.85)(0.43,0.07){1}{\line(1,0){0.43}}
\multiput(115.5,37.91)(0.42,0.07){1}{\line(1,0){0.42}}
\multiput(115.92,37.98)(0.4,0.08){1}{\line(1,0){0.4}}
\multiput(116.32,38.06)(0.38,0.08){1}{\line(1,0){0.38}}
\multiput(116.7,38.14)(0.37,0.09){1}{\line(1,0){0.37}}
\multiput(117.07,38.23)(0.35,0.09){1}{\line(1,0){0.35}}
\multiput(117.42,38.32)(0.33,0.1){1}{\line(1,0){0.33}}
\multiput(117.75,38.42)(0.31,0.1){1}{\line(1,0){0.31}}
\multiput(118.06,38.52)(0.29,0.1){1}{\line(1,0){0.29}}
\multiput(118.35,38.62)(0.27,0.11){1}{\line(1,0){0.27}}
\multiput(118.62,38.73)(0.25,0.11){1}{\line(1,0){0.25}}
\multiput(118.86,38.84)(0.22,0.11){1}{\line(1,0){0.22}}
\multiput(119.09,38.96)(0.2,0.12){1}{\line(1,0){0.2}}
\multiput(119.29,39.07)(0.18,0.12){1}{\line(1,0){0.18}}
\multiput(119.46,39.19)(0.15,0.12){1}{\line(1,0){0.15}}
\multiput(119.61,39.31)(0.13,0.12){1}{\line(1,0){0.13}}
\multiput(119.74,39.43)(0.1,0.12){1}{\line(0,1){0.12}}
\multiput(119.84,39.56)(0.08,0.13){1}{\line(0,1){0.13}}
\multiput(119.92,39.68)(0.05,0.13){1}{\line(0,1){0.13}}
\multiput(119.97,39.81)(0.03,0.13){1}{\line(0,1){0.13}}
\linethickness{0.3mm}
\put(90,39.94){\line(0,1){0.13}}
\multiput(89.97,40.19)(0.03,-0.13){1}{\line(0,-1){0.13}}
\multiput(89.92,40.32)(0.05,-0.13){1}{\line(0,-1){0.13}}
\multiput(89.84,40.44)(0.08,-0.13){1}{\line(0,-1){0.13}}
\multiput(89.74,40.57)(0.1,-0.12){1}{\line(0,-1){0.12}}
\multiput(89.61,40.69)(0.13,-0.12){1}{\line(1,0){0.13}}
\multiput(89.46,40.81)(0.15,-0.12){1}{\line(1,0){0.15}}
\multiput(89.29,40.93)(0.18,-0.12){1}{\line(1,0){0.18}}
\multiput(89.09,41.04)(0.2,-0.12){1}{\line(1,0){0.2}}
\multiput(88.86,41.16)(0.22,-0.11){1}{\line(1,0){0.22}}
\multiput(88.62,41.27)(0.25,-0.11){1}{\line(1,0){0.25}}
\multiput(88.35,41.38)(0.27,-0.11){1}{\line(1,0){0.27}}
\multiput(88.06,41.48)(0.29,-0.1){1}{\line(1,0){0.29}}
\multiput(87.75,41.58)(0.31,-0.1){1}{\line(1,0){0.31}}
\multiput(87.42,41.68)(0.33,-0.1){1}{\line(1,0){0.33}}
\multiput(87.07,41.77)(0.35,-0.09){1}{\line(1,0){0.35}}
\multiput(86.7,41.86)(0.37,-0.09){1}{\line(1,0){0.37}}
\multiput(86.32,41.94)(0.38,-0.08){1}{\line(1,0){0.38}}
\multiput(85.92,42.02)(0.4,-0.08){1}{\line(1,0){0.4}}
\multiput(85.5,42.09)(0.42,-0.07){1}{\line(1,0){0.42}}
\multiput(85.07,42.15)(0.43,-0.07){1}{\line(1,0){0.43}}
\multiput(84.63,42.22)(0.44,-0.06){1}{\line(1,0){0.44}}
\multiput(84.18,42.27)(0.45,-0.06){1}{\line(1,0){0.45}}
\multiput(83.71,42.32)(0.47,-0.05){1}{\line(1,0){0.47}}
\multiput(83.23,42.37)(0.48,-0.04){1}{\line(1,0){0.48}}
\multiput(82.75,42.4)(0.48,-0.04){1}{\line(1,0){0.48}}
\multiput(82.26,42.44)(0.49,-0.03){1}{\line(1,0){0.49}}
\multiput(81.76,42.46)(0.5,-0.03){1}{\line(1,0){0.5}}
\multiput(81.26,42.48)(0.5,-0.02){1}{\line(1,0){0.5}}
\multiput(80.76,42.49)(0.5,-0.01){1}{\line(1,0){0.5}}
\multiput(80.25,42.5)(0.51,-0.01){1}{\line(1,0){0.51}}
\put(79.75,42.5){\line(1,0){0.51}}
\multiput(79.24,42.49)(0.51,0.01){1}{\line(1,0){0.51}}
\multiput(78.74,42.48)(0.5,0.01){1}{\line(1,0){0.5}}
\multiput(78.24,42.46)(0.5,0.02){1}{\line(1,0){0.5}}
\multiput(77.74,42.44)(0.5,0.03){1}{\line(1,0){0.5}}
\multiput(77.25,42.4)(0.49,0.03){1}{\line(1,0){0.49}}
\multiput(76.77,42.37)(0.48,0.04){1}{\line(1,0){0.48}}
\multiput(76.29,42.32)(0.48,0.04){1}{\line(1,0){0.48}}
\multiput(75.82,42.27)(0.47,0.05){1}{\line(1,0){0.47}}
\multiput(75.37,42.22)(0.45,0.06){1}{\line(1,0){0.45}}
\multiput(74.93,42.15)(0.44,0.06){1}{\line(1,0){0.44}}
\multiput(74.5,42.09)(0.43,0.07){1}{\line(1,0){0.43}}
\multiput(74.08,42.02)(0.42,0.07){1}{\line(1,0){0.42}}
\multiput(73.68,41.94)(0.4,0.08){1}{\line(1,0){0.4}}
\multiput(73.3,41.86)(0.38,0.08){1}{\line(1,0){0.38}}
\multiput(72.93,41.77)(0.37,0.09){1}{\line(1,0){0.37}}
\multiput(72.58,41.68)(0.35,0.09){1}{\line(1,0){0.35}}
\multiput(72.25,41.58)(0.33,0.1){1}{\line(1,0){0.33}}
\multiput(71.94,41.48)(0.31,0.1){1}{\line(1,0){0.31}}
\multiput(71.65,41.38)(0.29,0.1){1}{\line(1,0){0.29}}
\multiput(71.38,41.27)(0.27,0.11){1}{\line(1,0){0.27}}
\multiput(71.14,41.16)(0.25,0.11){1}{\line(1,0){0.25}}
\multiput(70.91,41.04)(0.22,0.11){1}{\line(1,0){0.22}}
\multiput(70.71,40.93)(0.2,0.12){1}{\line(1,0){0.2}}
\multiput(70.54,40.81)(0.18,0.12){1}{\line(1,0){0.18}}
\multiput(70.39,40.69)(0.15,0.12){1}{\line(1,0){0.15}}
\multiput(70.26,40.57)(0.13,0.12){1}{\line(1,0){0.13}}
\multiput(70.16,40.44)(0.1,0.12){1}{\line(0,1){0.12}}
\multiput(70.08,40.32)(0.08,0.13){1}{\line(0,1){0.13}}
\multiput(70.03,40.19)(0.05,0.13){1}{\line(0,1){0.13}}
\multiput(70,40.06)(0.03,0.13){1}{\line(0,1){0.13}}
\put(70,39.94){\line(0,1){0.13}}
\multiput(70,39.94)(0.03,-0.13){1}{\line(0,-1){0.13}}
\multiput(70.03,39.81)(0.05,-0.13){1}{\line(0,-1){0.13}}
\multiput(70.08,39.68)(0.08,-0.13){1}{\line(0,-1){0.13}}
\multiput(70.16,39.56)(0.1,-0.12){1}{\line(0,-1){0.12}}
\multiput(70.26,39.43)(0.13,-0.12){1}{\line(1,0){0.13}}
\multiput(70.39,39.31)(0.15,-0.12){1}{\line(1,0){0.15}}
\multiput(70.54,39.19)(0.18,-0.12){1}{\line(1,0){0.18}}
\multiput(70.71,39.07)(0.2,-0.12){1}{\line(1,0){0.2}}
\multiput(70.91,38.96)(0.22,-0.11){1}{\line(1,0){0.22}}
\multiput(71.14,38.84)(0.25,-0.11){1}{\line(1,0){0.25}}
\multiput(71.38,38.73)(0.27,-0.11){1}{\line(1,0){0.27}}
\multiput(71.65,38.62)(0.29,-0.1){1}{\line(1,0){0.29}}
\multiput(71.94,38.52)(0.31,-0.1){1}{\line(1,0){0.31}}
\multiput(72.25,38.42)(0.33,-0.1){1}{\line(1,0){0.33}}
\multiput(72.58,38.32)(0.35,-0.09){1}{\line(1,0){0.35}}
\multiput(72.93,38.23)(0.37,-0.09){1}{\line(1,0){0.37}}
\multiput(73.3,38.14)(0.38,-0.08){1}{\line(1,0){0.38}}
\multiput(73.68,38.06)(0.4,-0.08){1}{\line(1,0){0.4}}
\multiput(74.08,37.98)(0.42,-0.07){1}{\line(1,0){0.42}}
\multiput(74.5,37.91)(0.43,-0.07){1}{\line(1,0){0.43}}
\multiput(74.93,37.85)(0.44,-0.06){1}{\line(1,0){0.44}}
\multiput(75.37,37.78)(0.45,-0.06){1}{\line(1,0){0.45}}
\multiput(75.82,37.73)(0.47,-0.05){1}{\line(1,0){0.47}}
\multiput(76.29,37.68)(0.48,-0.04){1}{\line(1,0){0.48}}
\multiput(76.77,37.63)(0.48,-0.04){1}{\line(1,0){0.48}}
\multiput(77.25,37.6)(0.49,-0.03){1}{\line(1,0){0.49}}
\multiput(77.74,37.56)(0.5,-0.03){1}{\line(1,0){0.5}}
\multiput(78.24,37.54)(0.5,-0.02){1}{\line(1,0){0.5}}
\multiput(78.74,37.52)(0.5,-0.01){1}{\line(1,0){0.5}}
\multiput(79.24,37.51)(0.51,-0.01){1}{\line(1,0){0.51}}
\put(79.75,37.5){\line(1,0){0.51}}
\multiput(80.25,37.5)(0.51,0.01){1}{\line(1,0){0.51}}
\multiput(80.76,37.51)(0.5,0.01){1}{\line(1,0){0.5}}
\multiput(81.26,37.52)(0.5,0.02){1}{\line(1,0){0.5}}
\multiput(81.76,37.54)(0.5,0.03){1}{\line(1,0){0.5}}
\multiput(82.26,37.56)(0.49,0.03){1}{\line(1,0){0.49}}
\multiput(82.75,37.6)(0.48,0.04){1}{\line(1,0){0.48}}
\multiput(83.23,37.63)(0.48,0.04){1}{\line(1,0){0.48}}
\multiput(83.71,37.68)(0.47,0.05){1}{\line(1,0){0.47}}
\multiput(84.18,37.73)(0.45,0.06){1}{\line(1,0){0.45}}
\multiput(84.63,37.78)(0.44,0.06){1}{\line(1,0){0.44}}
\multiput(85.07,37.85)(0.43,0.07){1}{\line(1,0){0.43}}
\multiput(85.5,37.91)(0.42,0.07){1}{\line(1,0){0.42}}
\multiput(85.92,37.98)(0.4,0.08){1}{\line(1,0){0.4}}
\multiput(86.32,38.06)(0.38,0.08){1}{\line(1,0){0.38}}
\multiput(86.7,38.14)(0.37,0.09){1}{\line(1,0){0.37}}
\multiput(87.07,38.23)(0.35,0.09){1}{\line(1,0){0.35}}
\multiput(87.42,38.32)(0.33,0.1){1}{\line(1,0){0.33}}
\multiput(87.75,38.42)(0.31,0.1){1}{\line(1,0){0.31}}
\multiput(88.06,38.52)(0.29,0.1){1}{\line(1,0){0.29}}
\multiput(88.35,38.62)(0.27,0.11){1}{\line(1,0){0.27}}
\multiput(88.62,38.73)(0.25,0.11){1}{\line(1,0){0.25}}
\multiput(88.86,38.84)(0.22,0.11){1}{\line(1,0){0.22}}
\multiput(89.09,38.96)(0.2,0.12){1}{\line(1,0){0.2}}
\multiput(89.29,39.07)(0.18,0.12){1}{\line(1,0){0.18}}
\multiput(89.46,39.19)(0.15,0.12){1}{\line(1,0){0.15}}
\multiput(89.61,39.31)(0.13,0.12){1}{\line(1,0){0.13}}
\multiput(89.74,39.43)(0.1,0.12){1}{\line(0,1){0.12}}
\multiput(89.84,39.56)(0.08,0.13){1}{\line(0,1){0.13}}
\multiput(89.92,39.68)(0.05,0.13){1}{\line(0,1){0.13}}
\multiput(89.97,39.81)(0.03,0.13){1}{\line(0,1){0.13}}
\put(5,40){\makebox(0,0)[cc]{+}}
\put(55,40){\makebox(0,0)[cc]{+}}
\put(137,40){\makebox(0,0)[cc]{+ $\cdots$}}
\put(30,40){\makebox(0,0)[cc]{1PI}}
\put(80,40){\makebox(0,0)[cc]{1PI}}
\put(110,40){\makebox(0,0)[cc]{1PI}}
\end{picture}
}

\def\figb{
\def\JPicScale{0.8}
\ifx\JPicScale\undefined\def\JPicScale{1}\fi
\unitlength \JPicScale mm
\begin{picture}(125,75)(0,0)
\linethickness{0.3mm}
\put(41.36,44.98){\line(0,1){0.49}}
\multiput(41.34,45.97)(0.02,-0.49){1}{\line(0,-1){0.49}}
\multiput(41.29,46.46)(0.05,-0.49){1}{\line(0,-1){0.49}}
\multiput(41.23,46.95)(0.07,-0.49){1}{\line(0,-1){0.49}}
\multiput(41.14,47.43)(0.09,-0.49){1}{\line(0,-1){0.49}}
\multiput(41.02,47.91)(0.11,-0.48){1}{\line(0,-1){0.48}}
\multiput(40.89,48.39)(0.14,-0.47){1}{\line(0,-1){0.47}}
\multiput(40.73,48.86)(0.16,-0.47){1}{\line(0,-1){0.47}}
\multiput(40.55,49.32)(0.18,-0.46){1}{\line(0,-1){0.46}}
\multiput(40.35,49.77)(0.1,-0.23){2}{\line(0,-1){0.23}}
\multiput(40.13,50.21)(0.11,-0.22){2}{\line(0,-1){0.22}}
\multiput(39.89,50.64)(0.12,-0.22){2}{\line(0,-1){0.22}}
\multiput(39.63,51.06)(0.13,-0.21){2}{\line(0,-1){0.21}}
\multiput(39.35,51.47)(0.14,-0.2){2}{\line(0,-1){0.2}}
\multiput(39.06,51.86)(0.15,-0.2){2}{\line(0,-1){0.2}}
\multiput(38.74,52.24)(0.11,-0.13){3}{\line(0,-1){0.13}}
\multiput(38.41,52.61)(0.11,-0.12){3}{\line(0,-1){0.12}}
\multiput(38.06,52.95)(0.12,-0.12){3}{\line(1,0){0.12}}
\multiput(37.7,53.29)(0.12,-0.11){3}{\line(1,0){0.12}}
\multiput(37.32,53.6)(0.13,-0.11){3}{\line(1,0){0.13}}
\multiput(36.92,53.9)(0.2,-0.15){2}{\line(1,0){0.2}}
\multiput(36.51,54.18)(0.2,-0.14){2}{\line(1,0){0.2}}
\multiput(36.1,54.44)(0.21,-0.13){2}{\line(1,0){0.21}}
\multiput(35.66,54.68)(0.22,-0.12){2}{\line(1,0){0.22}}
\multiput(35.22,54.9)(0.22,-0.11){2}{\line(1,0){0.22}}
\multiput(34.77,55.1)(0.23,-0.1){2}{\line(1,0){0.23}}
\multiput(34.31,55.28)(0.46,-0.18){1}{\line(1,0){0.46}}
\multiput(33.84,55.43)(0.47,-0.16){1}{\line(1,0){0.47}}
\multiput(33.37,55.57)(0.47,-0.14){1}{\line(1,0){0.47}}
\multiput(32.89,55.68)(0.48,-0.11){1}{\line(1,0){0.48}}
\multiput(32.4,55.77)(0.49,-0.09){1}{\line(1,0){0.49}}
\multiput(31.91,55.84)(0.49,-0.07){1}{\line(1,0){0.49}}
\multiput(31.42,55.89)(0.49,-0.05){1}{\line(1,0){0.49}}
\multiput(30.93,55.91)(0.49,-0.02){1}{\line(1,0){0.49}}
\put(30.44,55.91){\line(1,0){0.49}}
\multiput(29.94,55.89)(0.49,0.02){1}{\line(1,0){0.49}}
\multiput(29.45,55.84)(0.49,0.05){1}{\line(1,0){0.49}}
\multiput(28.96,55.77)(0.49,0.07){1}{\line(1,0){0.49}}
\multiput(28.48,55.68)(0.49,0.09){1}{\line(1,0){0.49}}
\multiput(28,55.57)(0.48,0.11){1}{\line(1,0){0.48}}
\multiput(27.52,55.43)(0.47,0.14){1}{\line(1,0){0.47}}
\multiput(27.05,55.28)(0.47,0.16){1}{\line(1,0){0.47}}
\multiput(26.59,55.1)(0.46,0.18){1}{\line(1,0){0.46}}
\multiput(26.14,54.9)(0.23,0.1){2}{\line(1,0){0.23}}
\multiput(25.7,54.68)(0.22,0.11){2}{\line(1,0){0.22}}
\multiput(25.27,54.44)(0.22,0.12){2}{\line(1,0){0.22}}
\multiput(24.85,54.18)(0.21,0.13){2}{\line(1,0){0.21}}
\multiput(24.44,53.9)(0.2,0.14){2}{\line(1,0){0.2}}
\multiput(24.05,53.6)(0.2,0.15){2}{\line(1,0){0.2}}
\multiput(23.67,53.29)(0.13,0.11){3}{\line(1,0){0.13}}
\multiput(23.3,52.95)(0.12,0.11){3}{\line(1,0){0.12}}
\multiput(22.95,52.61)(0.12,0.12){3}{\line(0,1){0.12}}
\multiput(22.62,52.24)(0.11,0.12){3}{\line(0,1){0.12}}
\multiput(22.31,51.86)(0.11,0.13){3}{\line(0,1){0.13}}
\multiput(22.01,51.47)(0.15,0.2){2}{\line(0,1){0.2}}
\multiput(21.73,51.06)(0.14,0.2){2}{\line(0,1){0.2}}
\multiput(21.47,50.64)(0.13,0.21){2}{\line(0,1){0.21}}
\multiput(21.23,50.21)(0.12,0.22){2}{\line(0,1){0.22}}
\multiput(21.01,49.77)(0.11,0.22){2}{\line(0,1){0.22}}
\multiput(20.81,49.32)(0.1,0.23){2}{\line(0,1){0.23}}
\multiput(20.63,48.86)(0.18,0.46){1}{\line(0,1){0.46}}
\multiput(20.48,48.39)(0.16,0.47){1}{\line(0,1){0.47}}
\multiput(20.34,47.91)(0.14,0.47){1}{\line(0,1){0.47}}
\multiput(20.23,47.43)(0.11,0.48){1}{\line(0,1){0.48}}
\multiput(20.14,46.95)(0.09,0.49){1}{\line(0,1){0.49}}
\multiput(20.07,46.46)(0.07,0.49){1}{\line(0,1){0.49}}
\multiput(20.02,45.97)(0.05,0.49){1}{\line(0,1){0.49}}
\multiput(20,45.47)(0.02,0.49){1}{\line(0,1){0.49}}
\put(20,44.98){\line(0,1){0.49}}
\multiput(20,44.98)(0.02,-0.49){1}{\line(0,-1){0.49}}
\multiput(20.02,44.49)(0.05,-0.49){1}{\line(0,-1){0.49}}
\multiput(20.07,44)(0.07,-0.49){1}{\line(0,-1){0.49}}
\multiput(20.14,43.51)(0.09,-0.49){1}{\line(0,-1){0.49}}
\multiput(20.23,43.02)(0.11,-0.48){1}{\line(0,-1){0.48}}
\multiput(20.34,42.54)(0.14,-0.47){1}{\line(0,-1){0.47}}
\multiput(20.48,42.07)(0.16,-0.47){1}{\line(0,-1){0.47}}
\multiput(20.63,41.6)(0.18,-0.46){1}{\line(0,-1){0.46}}
\multiput(20.81,41.14)(0.1,-0.23){2}{\line(0,-1){0.23}}
\multiput(21.01,40.69)(0.11,-0.22){2}{\line(0,-1){0.22}}
\multiput(21.23,40.25)(0.12,-0.22){2}{\line(0,-1){0.22}}
\multiput(21.47,39.81)(0.13,-0.21){2}{\line(0,-1){0.21}}
\multiput(21.73,39.39)(0.14,-0.2){2}{\line(0,-1){0.2}}
\multiput(22.01,38.99)(0.15,-0.2){2}{\line(0,-1){0.2}}
\multiput(22.31,38.59)(0.11,-0.13){3}{\line(0,-1){0.13}}
\multiput(22.62,38.21)(0.11,-0.12){3}{\line(0,-1){0.12}}
\multiput(22.95,37.85)(0.12,-0.12){3}{\line(0,-1){0.12}}
\multiput(23.3,37.5)(0.12,-0.11){3}{\line(1,0){0.12}}
\multiput(23.67,37.17)(0.13,-0.11){3}{\line(1,0){0.13}}
\multiput(24.05,36.85)(0.2,-0.15){2}{\line(1,0){0.2}}
\multiput(24.44,36.55)(0.2,-0.14){2}{\line(1,0){0.2}}
\multiput(24.85,36.28)(0.21,-0.13){2}{\line(1,0){0.21}}
\multiput(25.27,36.02)(0.22,-0.12){2}{\line(1,0){0.22}}
\multiput(25.7,35.78)(0.22,-0.11){2}{\line(1,0){0.22}}
\multiput(26.14,35.56)(0.23,-0.1){2}{\line(1,0){0.23}}
\multiput(26.59,35.36)(0.46,-0.18){1}{\line(1,0){0.46}}
\multiput(27.05,35.18)(0.47,-0.16){1}{\line(1,0){0.47}}
\multiput(27.52,35.02)(0.47,-0.14){1}{\line(1,0){0.47}}
\multiput(28,34.89)(0.48,-0.11){1}{\line(1,0){0.48}}
\multiput(28.48,34.77)(0.49,-0.09){1}{\line(1,0){0.49}}
\multiput(28.96,34.68)(0.49,-0.07){1}{\line(1,0){0.49}}
\multiput(29.45,34.61)(0.49,-0.05){1}{\line(1,0){0.49}}
\multiput(29.94,34.57)(0.49,-0.02){1}{\line(1,0){0.49}}
\put(30.44,34.55){\line(1,0){0.49}}
\multiput(30.93,34.55)(0.49,0.02){1}{\line(1,0){0.49}}
\multiput(31.42,34.57)(0.49,0.05){1}{\line(1,0){0.49}}
\multiput(31.91,34.61)(0.49,0.07){1}{\line(1,0){0.49}}
\multiput(32.4,34.68)(0.49,0.09){1}{\line(1,0){0.49}}
\multiput(32.89,34.77)(0.48,0.11){1}{\line(1,0){0.48}}
\multiput(33.37,34.89)(0.47,0.14){1}{\line(1,0){0.47}}
\multiput(33.84,35.02)(0.47,0.16){1}{\line(1,0){0.47}}
\multiput(34.31,35.18)(0.46,0.18){1}{\line(1,0){0.46}}
\multiput(34.77,35.36)(0.23,0.1){2}{\line(1,0){0.23}}
\multiput(35.22,35.56)(0.22,0.11){2}{\line(1,0){0.22}}
\multiput(35.66,35.78)(0.22,0.12){2}{\line(1,0){0.22}}
\multiput(36.1,36.02)(0.21,0.13){2}{\line(1,0){0.21}}
\multiput(36.51,36.28)(0.2,0.14){2}{\line(1,0){0.2}}
\multiput(36.92,36.55)(0.2,0.15){2}{\line(1,0){0.2}}
\multiput(37.32,36.85)(0.13,0.11){3}{\line(1,0){0.13}}
\multiput(37.7,37.17)(0.12,0.11){3}{\line(1,0){0.12}}
\multiput(38.06,37.5)(0.12,0.12){3}{\line(0,1){0.12}}
\multiput(38.41,37.85)(0.11,0.12){3}{\line(0,1){0.12}}
\multiput(38.74,38.21)(0.11,0.13){3}{\line(0,1){0.13}}
\multiput(39.06,38.59)(0.15,0.2){2}{\line(0,1){0.2}}
\multiput(39.35,38.99)(0.14,0.2){2}{\line(0,1){0.2}}
\multiput(39.63,39.39)(0.13,0.21){2}{\line(0,1){0.21}}
\multiput(39.89,39.81)(0.12,0.22){2}{\line(0,1){0.22}}
\multiput(40.13,40.25)(0.11,0.22){2}{\line(0,1){0.22}}
\multiput(40.35,40.69)(0.1,0.23){2}{\line(0,1){0.23}}
\multiput(40.55,41.14)(0.18,0.46){1}{\line(0,1){0.46}}
\multiput(40.73,41.6)(0.16,0.47){1}{\line(0,1){0.47}}
\multiput(40.89,42.07)(0.14,0.47){1}{\line(0,1){0.47}}
\multiput(41.02,42.54)(0.11,0.48){1}{\line(0,1){0.48}}
\multiput(41.14,43.02)(0.09,0.49){1}{\line(0,1){0.49}}
\multiput(41.23,43.51)(0.07,0.49){1}{\line(0,1){0.49}}
\multiput(41.29,44)(0.05,0.49){1}{\line(0,1){0.49}}
\multiput(41.34,44.49)(0.02,0.49){1}{\line(0,1){0.49}}
\linethickness{0.3mm}
\put(101.37,45.44){\line(0,1){0.49}}
\multiput(101.34,46.42)(0.02,-0.49){1}{\line(0,-1){0.49}}
\multiput(101.3,46.92)(0.05,-0.49){1}{\line(0,-1){0.49}}
\multiput(101.23,47.4)(0.07,-0.49){1}{\line(0,-1){0.49}}
\multiput(101.14,47.89)(0.09,-0.49){1}{\line(0,-1){0.49}}
\multiput(101.03,48.37)(0.11,-0.48){1}{\line(0,-1){0.48}}
\multiput(100.89,48.84)(0.14,-0.47){1}{\line(0,-1){0.47}}
\multiput(100.73,49.31)(0.16,-0.47){1}{\line(0,-1){0.47}}
\multiput(100.56,49.77)(0.18,-0.46){1}{\line(0,-1){0.46}}
\multiput(100.36,50.22)(0.1,-0.23){2}{\line(0,-1){0.23}}
\multiput(100.14,50.67)(0.11,-0.22){2}{\line(0,-1){0.22}}
\multiput(99.9,51.1)(0.12,-0.22){2}{\line(0,-1){0.22}}
\multiput(99.64,51.52)(0.13,-0.21){2}{\line(0,-1){0.21}}
\multiput(99.36,51.92)(0.14,-0.2){2}{\line(0,-1){0.2}}
\multiput(99.06,52.32)(0.15,-0.2){2}{\line(0,-1){0.2}}
\multiput(98.74,52.7)(0.11,-0.13){3}{\line(0,-1){0.13}}
\multiput(98.41,53.06)(0.11,-0.12){3}{\line(0,-1){0.12}}
\multiput(98.06,53.41)(0.12,-0.12){3}{\line(1,0){0.12}}
\multiput(97.7,53.74)(0.12,-0.11){3}{\line(1,0){0.12}}
\multiput(97.32,54.06)(0.13,-0.11){3}{\line(1,0){0.13}}
\multiput(96.92,54.36)(0.2,-0.15){2}{\line(1,0){0.2}}
\multiput(96.52,54.64)(0.2,-0.14){2}{\line(1,0){0.2}}
\multiput(96.1,54.9)(0.21,-0.13){2}{\line(1,0){0.21}}
\multiput(95.67,55.14)(0.22,-0.12){2}{\line(1,0){0.22}}
\multiput(95.22,55.36)(0.22,-0.11){2}{\line(1,0){0.22}}
\multiput(94.77,55.56)(0.23,-0.1){2}{\line(1,0){0.23}}
\multiput(94.31,55.73)(0.46,-0.18){1}{\line(1,0){0.46}}
\multiput(93.84,55.89)(0.47,-0.16){1}{\line(1,0){0.47}}
\multiput(93.37,56.03)(0.47,-0.14){1}{\line(1,0){0.47}}
\multiput(92.89,56.14)(0.48,-0.11){1}{\line(1,0){0.48}}
\multiput(92.4,56.23)(0.49,-0.09){1}{\line(1,0){0.49}}
\multiput(91.92,56.3)(0.49,-0.07){1}{\line(1,0){0.49}}
\multiput(91.42,56.34)(0.49,-0.05){1}{\line(1,0){0.49}}
\multiput(90.93,56.37)(0.49,-0.02){1}{\line(1,0){0.49}}
\put(90.44,56.37){\line(1,0){0.49}}
\multiput(89.94,56.34)(0.49,0.02){1}{\line(1,0){0.49}}
\multiput(89.45,56.3)(0.49,0.05){1}{\line(1,0){0.49}}
\multiput(88.96,56.23)(0.49,0.07){1}{\line(1,0){0.49}}
\multiput(88.48,56.14)(0.49,0.09){1}{\line(1,0){0.49}}
\multiput(88,56.03)(0.48,0.11){1}{\line(1,0){0.48}}
\multiput(87.52,55.89)(0.47,0.14){1}{\line(1,0){0.47}}
\multiput(87.06,55.73)(0.47,0.16){1}{\line(1,0){0.47}}
\multiput(86.6,55.56)(0.46,0.18){1}{\line(1,0){0.46}}
\multiput(86.14,55.36)(0.23,0.1){2}{\line(1,0){0.23}}
\multiput(85.7,55.14)(0.22,0.11){2}{\line(1,0){0.22}}
\multiput(85.27,54.9)(0.22,0.12){2}{\line(1,0){0.22}}
\multiput(84.85,54.64)(0.21,0.13){2}{\line(1,0){0.21}}
\multiput(84.44,54.36)(0.2,0.14){2}{\line(1,0){0.2}}
\multiput(84.05,54.06)(0.2,0.15){2}{\line(1,0){0.2}}
\multiput(83.67,53.74)(0.13,0.11){3}{\line(1,0){0.13}}
\multiput(83.31,53.41)(0.12,0.11){3}{\line(1,0){0.12}}
\multiput(82.96,53.06)(0.12,0.12){3}{\line(1,0){0.12}}
\multiput(82.62,52.7)(0.11,0.12){3}{\line(0,1){0.12}}
\multiput(82.31,52.32)(0.11,0.13){3}{\line(0,1){0.13}}
\multiput(82.01,51.92)(0.15,0.2){2}{\line(0,1){0.2}}
\multiput(81.73,51.52)(0.14,0.2){2}{\line(0,1){0.2}}
\multiput(81.47,51.1)(0.13,0.21){2}{\line(0,1){0.21}}
\multiput(81.23,50.67)(0.12,0.22){2}{\line(0,1){0.22}}
\multiput(81.01,50.22)(0.11,0.22){2}{\line(0,1){0.22}}
\multiput(80.81,49.77)(0.1,0.23){2}{\line(0,1){0.23}}
\multiput(80.63,49.31)(0.18,0.46){1}{\line(0,1){0.46}}
\multiput(80.48,48.84)(0.16,0.47){1}{\line(0,1){0.47}}
\multiput(80.34,48.37)(0.14,0.47){1}{\line(0,1){0.47}}
\multiput(80.23,47.89)(0.11,0.48){1}{\line(0,1){0.48}}
\multiput(80.14,47.4)(0.09,0.49){1}{\line(0,1){0.49}}
\multiput(80.07,46.92)(0.07,0.49){1}{\line(0,1){0.49}}
\multiput(80.03,46.42)(0.05,0.49){1}{\line(0,1){0.49}}
\multiput(80,45.93)(0.02,0.49){1}{\line(0,1){0.49}}
\put(80,45.44){\line(0,1){0.49}}
\multiput(80,45.44)(0.02,-0.49){1}{\line(0,-1){0.49}}
\multiput(80.03,44.94)(0.05,-0.49){1}{\line(0,-1){0.49}}
\multiput(80.07,44.45)(0.07,-0.49){1}{\line(0,-1){0.49}}
\multiput(80.14,43.96)(0.09,-0.49){1}{\line(0,-1){0.49}}
\multiput(80.23,43.48)(0.11,-0.48){1}{\line(0,-1){0.48}}
\multiput(80.34,43)(0.14,-0.47){1}{\line(0,-1){0.47}}
\multiput(80.48,42.52)(0.16,-0.47){1}{\line(0,-1){0.47}}
\multiput(80.63,42.06)(0.18,-0.46){1}{\line(0,-1){0.46}}
\multiput(80.81,41.6)(0.1,-0.23){2}{\line(0,-1){0.23}}
\multiput(81.01,41.14)(0.11,-0.22){2}{\line(0,-1){0.22}}
\multiput(81.23,40.7)(0.12,-0.22){2}{\line(0,-1){0.22}}
\multiput(81.47,40.27)(0.13,-0.21){2}{\line(0,-1){0.21}}
\multiput(81.73,39.85)(0.14,-0.2){2}{\line(0,-1){0.2}}
\multiput(82.01,39.44)(0.15,-0.2){2}{\line(0,-1){0.2}}
\multiput(82.31,39.05)(0.11,-0.13){3}{\line(0,-1){0.13}}
\multiput(82.62,38.67)(0.11,-0.12){3}{\line(0,-1){0.12}}
\multiput(82.96,38.31)(0.12,-0.12){3}{\line(1,0){0.12}}
\multiput(83.31,37.96)(0.12,-0.11){3}{\line(1,0){0.12}}
\multiput(83.67,37.62)(0.13,-0.11){3}{\line(1,0){0.13}}
\multiput(84.05,37.31)(0.2,-0.15){2}{\line(1,0){0.2}}
\multiput(84.44,37.01)(0.2,-0.14){2}{\line(1,0){0.2}}
\multiput(84.85,36.73)(0.21,-0.13){2}{\line(1,0){0.21}}
\multiput(85.27,36.47)(0.22,-0.12){2}{\line(1,0){0.22}}
\multiput(85.7,36.23)(0.22,-0.11){2}{\line(1,0){0.22}}
\multiput(86.14,36.01)(0.23,-0.1){2}{\line(1,0){0.23}}
\multiput(86.6,35.81)(0.46,-0.18){1}{\line(1,0){0.46}}
\multiput(87.06,35.63)(0.47,-0.16){1}{\line(1,0){0.47}}
\multiput(87.52,35.48)(0.47,-0.14){1}{\line(1,0){0.47}}
\multiput(88,35.34)(0.48,-0.11){1}{\line(1,0){0.48}}
\multiput(88.48,35.23)(0.49,-0.09){1}{\line(1,0){0.49}}
\multiput(88.96,35.14)(0.49,-0.07){1}{\line(1,0){0.49}}
\multiput(89.45,35.07)(0.49,-0.05){1}{\line(1,0){0.49}}
\multiput(89.94,35.03)(0.49,-0.02){1}{\line(1,0){0.49}}
\put(90.44,35){\line(1,0){0.49}}
\multiput(90.93,35)(0.49,0.02){1}{\line(1,0){0.49}}
\multiput(91.42,35.03)(0.49,0.05){1}{\line(1,0){0.49}}
\multiput(91.92,35.07)(0.49,0.07){1}{\line(1,0){0.49}}
\multiput(92.4,35.14)(0.49,0.09){1}{\line(1,0){0.49}}
\multiput(92.89,35.23)(0.48,0.11){1}{\line(1,0){0.48}}
\multiput(93.37,35.34)(0.47,0.14){1}{\line(1,0){0.47}}
\multiput(93.84,35.48)(0.47,0.16){1}{\line(1,0){0.47}}
\multiput(94.31,35.63)(0.46,0.18){1}{\line(1,0){0.46}}
\multiput(94.77,35.81)(0.23,0.1){2}{\line(1,0){0.23}}
\multiput(95.22,36.01)(0.22,0.11){2}{\line(1,0){0.22}}
\multiput(95.67,36.23)(0.22,0.12){2}{\line(1,0){0.22}}
\multiput(96.1,36.47)(0.21,0.13){2}{\line(1,0){0.21}}
\multiput(96.52,36.73)(0.2,0.14){2}{\line(1,0){0.2}}
\multiput(96.92,37.01)(0.2,0.15){2}{\line(1,0){0.2}}
\multiput(97.32,37.31)(0.13,0.11){3}{\line(1,0){0.13}}
\multiput(97.7,37.62)(0.12,0.11){3}{\line(1,0){0.12}}
\multiput(98.06,37.96)(0.12,0.12){3}{\line(0,1){0.12}}
\multiput(98.41,38.31)(0.11,0.12){3}{\line(0,1){0.12}}
\multiput(98.74,38.67)(0.11,0.13){3}{\line(0,1){0.13}}
\multiput(99.06,39.05)(0.15,0.2){2}{\line(0,1){0.2}}
\multiput(99.36,39.44)(0.14,0.2){2}{\line(0,1){0.2}}
\multiput(99.64,39.85)(0.13,0.21){2}{\line(0,1){0.21}}
\multiput(99.9,40.27)(0.12,0.22){2}{\line(0,1){0.22}}
\multiput(100.14,40.7)(0.11,0.22){2}{\line(0,1){0.22}}
\multiput(100.36,41.14)(0.1,0.23){2}{\line(0,1){0.23}}
\multiput(100.56,41.6)(0.18,0.46){1}{\line(0,1){0.46}}
\multiput(100.73,42.06)(0.16,0.47){1}{\line(0,1){0.47}}
\multiput(100.89,42.52)(0.14,0.47){1}{\line(0,1){0.47}}
\multiput(101.03,43)(0.11,0.48){1}{\line(0,1){0.48}}
\multiput(101.14,43.48)(0.09,0.49){1}{\line(0,1){0.49}}
\multiput(101.23,43.96)(0.07,0.49){1}{\line(0,1){0.49}}
\multiput(101.3,44.45)(0.05,0.49){1}{\line(0,1){0.49}}
\multiput(101.34,44.94)(0.02,0.49){1}{\line(0,1){0.49}}
\linethickness{0.3mm}
\put(70,44.87){\line(0,1){0.25}}
\multiput(69.97,45.38)(0.03,-0.25){1}{\line(0,-1){0.25}}
\multiput(69.92,45.63)(0.05,-0.25){1}{\line(0,-1){0.25}}
\multiput(69.84,45.88)(0.08,-0.25){1}{\line(0,-1){0.25}}
\multiput(69.74,46.13)(0.1,-0.25){1}{\line(0,-1){0.25}}
\multiput(69.61,46.38)(0.13,-0.25){1}{\line(0,-1){0.25}}
\multiput(69.46,46.62)(0.15,-0.24){1}{\line(0,-1){0.24}}
\multiput(69.29,46.85)(0.18,-0.24){1}{\line(0,-1){0.24}}
\multiput(69.09,47.09)(0.1,-0.12){2}{\line(0,-1){0.12}}
\multiput(68.86,47.31)(0.11,-0.11){2}{\line(0,-1){0.11}}
\multiput(68.62,47.54)(0.12,-0.11){2}{\line(1,0){0.12}}
\multiput(68.35,47.75)(0.13,-0.11){2}{\line(1,0){0.13}}
\multiput(68.06,47.96)(0.14,-0.1){2}{\line(1,0){0.14}}
\multiput(67.75,48.16)(0.16,-0.1){2}{\line(1,0){0.16}}
\multiput(67.42,48.35)(0.17,-0.1){2}{\line(1,0){0.17}}
\multiput(67.07,48.54)(0.17,-0.09){2}{\line(1,0){0.17}}
\multiput(66.7,48.71)(0.37,-0.17){1}{\line(1,0){0.37}}
\multiput(66.32,48.88)(0.38,-0.17){1}{\line(1,0){0.38}}
\multiput(65.92,49.03)(0.4,-0.16){1}{\line(1,0){0.4}}
\multiput(65.5,49.17)(0.42,-0.14){1}{\line(1,0){0.42}}
\multiput(65.07,49.31)(0.43,-0.13){1}{\line(1,0){0.43}}
\multiput(64.63,49.43)(0.44,-0.12){1}{\line(1,0){0.44}}
\multiput(64.18,49.54)(0.45,-0.11){1}{\line(1,0){0.45}}
\multiput(63.71,49.64)(0.47,-0.1){1}{\line(1,0){0.47}}
\multiput(63.23,49.73)(0.48,-0.09){1}{\line(1,0){0.48}}
\multiput(62.75,49.81)(0.48,-0.08){1}{\line(1,0){0.48}}
\multiput(62.26,49.87)(0.49,-0.06){1}{\line(1,0){0.49}}
\multiput(61.76,49.92)(0.5,-0.05){1}{\line(1,0){0.5}}
\multiput(61.26,49.96)(0.5,-0.04){1}{\line(1,0){0.5}}
\multiput(60.76,49.99)(0.5,-0.03){1}{\line(1,0){0.5}}
\multiput(60.25,50)(0.51,-0.01){1}{\line(1,0){0.51}}
\put(59.75,50){\line(1,0){0.51}}
\multiput(59.24,49.99)(0.51,0.01){1}{\line(1,0){0.51}}
\multiput(58.74,49.96)(0.5,0.03){1}{\line(1,0){0.5}}
\multiput(58.24,49.92)(0.5,0.04){1}{\line(1,0){0.5}}
\multiput(57.74,49.87)(0.5,0.05){1}{\line(1,0){0.5}}
\multiput(57.25,49.81)(0.49,0.06){1}{\line(1,0){0.49}}
\multiput(56.77,49.73)(0.48,0.08){1}{\line(1,0){0.48}}
\multiput(56.29,49.64)(0.48,0.09){1}{\line(1,0){0.48}}
\multiput(55.82,49.54)(0.47,0.1){1}{\line(1,0){0.47}}
\multiput(55.37,49.43)(0.45,0.11){1}{\line(1,0){0.45}}
\multiput(54.93,49.31)(0.44,0.12){1}{\line(1,0){0.44}}
\multiput(54.5,49.17)(0.43,0.13){1}{\line(1,0){0.43}}
\multiput(54.08,49.03)(0.42,0.14){1}{\line(1,0){0.42}}
\multiput(53.68,48.88)(0.4,0.16){1}{\line(1,0){0.4}}
\multiput(53.3,48.71)(0.38,0.17){1}{\line(1,0){0.38}}
\multiput(52.93,48.54)(0.37,0.17){1}{\line(1,0){0.37}}
\multiput(52.58,48.35)(0.17,0.09){2}{\line(1,0){0.17}}
\multiput(52.25,48.16)(0.17,0.1){2}{\line(1,0){0.17}}
\multiput(51.94,47.96)(0.16,0.1){2}{\line(1,0){0.16}}
\multiput(51.65,47.75)(0.14,0.1){2}{\line(1,0){0.14}}
\multiput(51.38,47.54)(0.13,0.11){2}{\line(1,0){0.13}}
\multiput(51.14,47.31)(0.12,0.11){2}{\line(1,0){0.12}}
\multiput(50.91,47.09)(0.11,0.11){2}{\line(0,1){0.11}}
\multiput(50.71,46.85)(0.1,0.12){2}{\line(0,1){0.12}}
\multiput(50.54,46.62)(0.18,0.24){1}{\line(0,1){0.24}}
\multiput(50.39,46.38)(0.15,0.24){1}{\line(0,1){0.24}}
\multiput(50.26,46.13)(0.13,0.25){1}{\line(0,1){0.25}}
\multiput(50.16,45.88)(0.1,0.25){1}{\line(0,1){0.25}}
\multiput(50.08,45.63)(0.08,0.25){1}{\line(0,1){0.25}}
\multiput(50.03,45.38)(0.05,0.25){1}{\line(0,1){0.25}}
\multiput(50,45.13)(0.03,0.25){1}{\line(0,1){0.25}}
\put(50,44.87){\line(0,1){0.25}}
\multiput(50,44.87)(0.03,-0.25){1}{\line(0,-1){0.25}}
\multiput(50.03,44.62)(0.05,-0.25){1}{\line(0,-1){0.25}}
\multiput(50.08,44.37)(0.08,-0.25){1}{\line(0,-1){0.25}}
\multiput(50.16,44.12)(0.1,-0.25){1}{\line(0,-1){0.25}}
\multiput(50.26,43.87)(0.13,-0.25){1}{\line(0,-1){0.25}}
\multiput(50.39,43.62)(0.15,-0.24){1}{\line(0,-1){0.24}}
\multiput(50.54,43.38)(0.18,-0.24){1}{\line(0,-1){0.24}}
\multiput(50.71,43.15)(0.1,-0.12){2}{\line(0,-1){0.12}}
\multiput(50.91,42.91)(0.11,-0.11){2}{\line(0,-1){0.11}}
\multiput(51.14,42.69)(0.12,-0.11){2}{\line(1,0){0.12}}
\multiput(51.38,42.46)(0.13,-0.11){2}{\line(1,0){0.13}}
\multiput(51.65,42.25)(0.14,-0.1){2}{\line(1,0){0.14}}
\multiput(51.94,42.04)(0.16,-0.1){2}{\line(1,0){0.16}}
\multiput(52.25,41.84)(0.17,-0.1){2}{\line(1,0){0.17}}
\multiput(52.58,41.65)(0.17,-0.09){2}{\line(1,0){0.17}}
\multiput(52.93,41.46)(0.37,-0.17){1}{\line(1,0){0.37}}
\multiput(53.3,41.29)(0.38,-0.17){1}{\line(1,0){0.38}}
\multiput(53.68,41.12)(0.4,-0.16){1}{\line(1,0){0.4}}
\multiput(54.08,40.97)(0.42,-0.14){1}{\line(1,0){0.42}}
\multiput(54.5,40.83)(0.43,-0.13){1}{\line(1,0){0.43}}
\multiput(54.93,40.69)(0.44,-0.12){1}{\line(1,0){0.44}}
\multiput(55.37,40.57)(0.45,-0.11){1}{\line(1,0){0.45}}
\multiput(55.82,40.46)(0.47,-0.1){1}{\line(1,0){0.47}}
\multiput(56.29,40.36)(0.48,-0.09){1}{\line(1,0){0.48}}
\multiput(56.77,40.27)(0.48,-0.08){1}{\line(1,0){0.48}}
\multiput(57.25,40.19)(0.49,-0.06){1}{\line(1,0){0.49}}
\multiput(57.74,40.13)(0.5,-0.05){1}{\line(1,0){0.5}}
\multiput(58.24,40.08)(0.5,-0.04){1}{\line(1,0){0.5}}
\multiput(58.74,40.04)(0.5,-0.03){1}{\line(1,0){0.5}}
\multiput(59.24,40.01)(0.51,-0.01){1}{\line(1,0){0.51}}
\put(59.75,40){\line(1,0){0.51}}
\multiput(60.25,40)(0.51,0.01){1}{\line(1,0){0.51}}
\multiput(60.76,40.01)(0.5,0.03){1}{\line(1,0){0.5}}
\multiput(61.26,40.04)(0.5,0.04){1}{\line(1,0){0.5}}
\multiput(61.76,40.08)(0.5,0.05){1}{\line(1,0){0.5}}
\multiput(62.26,40.13)(0.49,0.06){1}{\line(1,0){0.49}}
\multiput(62.75,40.19)(0.48,0.08){1}{\line(1,0){0.48}}
\multiput(63.23,40.27)(0.48,0.09){1}{\line(1,0){0.48}}
\multiput(63.71,40.36)(0.47,0.1){1}{\line(1,0){0.47}}
\multiput(64.18,40.46)(0.45,0.11){1}{\line(1,0){0.45}}
\multiput(64.63,40.57)(0.44,0.12){1}{\line(1,0){0.44}}
\multiput(65.07,40.69)(0.43,0.13){1}{\line(1,0){0.43}}
\multiput(65.5,40.83)(0.42,0.14){1}{\line(1,0){0.42}}
\multiput(65.92,40.97)(0.4,0.16){1}{\line(1,0){0.4}}
\multiput(66.32,41.12)(0.38,0.17){1}{\line(1,0){0.38}}
\multiput(66.7,41.29)(0.37,0.17){1}{\line(1,0){0.37}}
\multiput(67.07,41.46)(0.17,0.09){2}{\line(1,0){0.17}}
\multiput(67.42,41.65)(0.17,0.1){2}{\line(1,0){0.17}}
\multiput(67.75,41.84)(0.16,0.1){2}{\line(1,0){0.16}}
\multiput(68.06,42.04)(0.14,0.1){2}{\line(1,0){0.14}}
\multiput(68.35,42.25)(0.13,0.11){2}{\line(1,0){0.13}}
\multiput(68.62,42.46)(0.12,0.11){2}{\line(1,0){0.12}}
\multiput(68.86,42.69)(0.11,0.11){2}{\line(0,1){0.11}}
\multiput(69.09,42.91)(0.1,0.12){2}{\line(0,1){0.12}}
\multiput(69.29,43.15)(0.18,0.24){1}{\line(0,1){0.24}}
\multiput(69.46,43.38)(0.15,0.24){1}{\line(0,1){0.24}}
\multiput(69.61,43.62)(0.13,0.25){1}{\line(0,1){0.25}}
\multiput(69.74,43.87)(0.1,0.25){1}{\line(0,1){0.25}}
\multiput(69.84,44.12)(0.08,0.25){1}{\line(0,1){0.25}}
\multiput(69.92,44.37)(0.05,0.25){1}{\line(0,1){0.25}}
\multiput(69.97,44.62)(0.03,0.25){1}{\line(0,1){0.25}}
\linethickness{0.3mm}
\put(41,45){\line(1,0){9}}
\linethickness{0.3mm}
\put(70,45){\line(1,0){10}}
\linethickness{0.3mm}
\multiput(5,69)(0.16,-0.12){125}{\line(1,0){0.16}}
\linethickness{0.3mm}
\multiput(5,50)(0.36,-0.12){42}{\line(1,0){0.36}}
\linethickness{0.3mm}
\linethickness{0.3mm}
\multiput(12,15)(0.12,0.16){125}{\line(0,1){0.16}}
\linethickness{0.3mm}
\multiput(95,55)(0.12,0.12){170}{\line(1,0){0.12}}
\linethickness{0.3mm}
\multiput(101,45)(0.3,0.12){83}{\line(1,0){0.3}}
\linethickness{0.3mm}
\multiput(95,36)(0.2,-0.12){125}{\line(1,0){0.2}}
\put(58,45){\makebox(0,0)[cc]{Full}}
\put(75,48){\makebox(0,0)[cc]{$k$}}
\put(88,45){\makebox(0,0)[cc]{1PI}}
\put(28,45){\makebox(0,0)[cc]{1PI}}
\put(1,75){\makebox(0,0)[cc]{$k_1$}}
\put(5,55){\makebox(0,0)[cc]{$k_2$}}
\put(10,10){\makebox(0,0)[cc]{$k_m$}}
\put(120,75){\makebox(0,0)[cc]{$\ell_1$}}
\put(120,49){\makebox(0,0)[cc]{$\ell_2$}}
\put(120,25){\makebox(0,0)[cc]{$\ell_n$}}
\end{picture}
}

\begin{document}

\begin{flushright}
DAMTP-2014-1\\
HRI/ST/1401
\end{flushright}

\vskip 12pt

\baselineskip 24pt

\begin{center}
{\Large \bf  Mass Renormalization in String Theory: General States}

\end{center}

\vskip .6cm
\medskip

\vspace*{4.0ex}

\baselineskip=18pt

\centerline{\large \rm Roji Pius$^a$, Arnab Rudra$^b$ and Ashoke Sen$^a$}

\vspace*{4.0ex}

\centerline{\large \it ~$^a$Harish-Chandra Research Institute}
\centerline{\large \it  Chhatnag Road, Jhusi,
Allahabad 211019, India}
\centerline{\large \it ~$^b$Department of Applied Mathematics and Theoretical Physics}
\centerline{\large \it Wilberforce Road, Cambridge CB3 0WA, UK}

\vspace*{1.0ex}
\centerline{\small E-mail:  rojipius@mri.ernet.in, A.Rudra@damtp.cam.ac.uk, sen@mri.ernet.in}

\vspace*{5.0ex}

\centerline{\bf Abstract} \bigskip

In a previous paper we described a procedure for computing the renormalized 
masses and S-matrix elements in bosonic string
theory for a special class of massive states which do not mix with unphysical states under
renormalization. In this paper we extend this result to general states in bosonic string
theory, and argue that only the squares of renormalized physical masses appear as
the locations of the poles of the S-matrix of other physical states. 
We also discuss generalizations to Neveu-Schwarz sector states in heterotic
and superstring theories.

\vfill \eject

\baselineskip=18pt

\tableofcontents

\sectiono{Introduction} \label{sintro}

We now have a well defined algorithm for computing perturbative
S-matrix elements of massless gauge
particles and BPS states in string theory to all orders in perturbation 
theory\cite{1209.5461,Belopolsky,dp,Witten,1304.7798}. These
states have the property that their masses are not renormalized away from the tree level
values due to various underlying symmetries. However string theory also contains stable
and unstable particles whose masses are not protected from quantum corrections, and
a direct systematic computation of  the renormalized  masses and S-matrix elements of
these states is plagued with difficulties\cite{Weinberg,Seiberg,
OoguriSakai,Yamamoto,AS,Das,Rey,Sundborg,Marcus,Amano,
Lee,Berera,0812.3129,0903.3979}. 
The main difficulty arises from the fact that world-sheet conformal invariance requires us to
use vertex operators of dimension (0,0) for defining string amplitudes, and this 
condition on the dimension of the operator translates to requiring the momenta to satisfy the
tree level mass-shell condition. Thus in the presence of a mass renormalization we run into
an apparent conflict between the requirement of world-sheet conformal invariance and
renormalized mass-shell condition. 

In a previous paper\cite{1311.1257} we described a systematic
procedure for computing the renormalized masses and S-matrix elements of a special
class of states in bosonic string theory which do not mix with unphysical
states under renormalization. Our goal in this paper will be to generalize this 
procedure to general states in bosonic string theory. We shall also briefly
discuss extensions
to the Neveu-Schwarz (NS) sector states in superstring and heterotic string theories.

We shall now summarize the contents of the rest of the sections. 
The reason that we had to restrict our analysis to a special class of states in
\cite{1311.1257} was to avoid the mixing between physical and unphysical states
which are degenerate at tree level. In \S\ref{stoy} we construct an example of a
gauge theory
where the tree level spectrum in a particular gauge has accidental degeneracy between
physical and unphysical states. We then develop an algorithm for extracting the 
quantum corrected physical mass in this theory, with the aim of generalizing this to
string theory later. 

In \S\ref{sorganize} we review some basic results for on-shell states in closed
bosonic string theory,
dividing them into physical, unphysical and pure gauge states and discuss their
off-shell generalization. We also review  the prescription 
for defining off-shell amplitudes in string theory which depend on the choice of local
coordinates at the punctures where the vertex operators are inserted. Finally we discuss
the constraints imposed on the choice of local coordinate system 
from the requirement that they
be compatible with the plumbing fixture procedure for gluing two Riemann surfaces to
form a third one. This allows us to express an off-shell amplitude as sums of products of 
one particle irreducible contributions and propagators.

\S\ref{sphysical}-\S\ref{sall} contains our main results. In \S\ref{sphysical} we
generalize the method of \S\ref{stoy} for systematically computing the renormalized 
physical masses in string theory. We also show that at one loop order the
renormalized physical masses are independent of the choice of local coordinate
system but the renormalized masses in the unphysical / pure gauge sector do
depend on the choice of local coordinates. In \S\ref{spoles} we examine the
locations of the poles in the scattering amplitudes of
external massless / BPS / special states in the complex $-k^2$ plane where $k$ is
given by the sum of some specific subset of external momenta. We
find that the possible locations of the poles are precisely at the squares of
physical and unphysical
masses found using the general algorithm of \S\ref{sphysical}. We also show that at the
leading order the residues at the physical poles are non-vanishing in general but the
residues at the poles associated with 
the unphysical / pure gauge sector states vanish. In \S\ref{sall} we
combine the results of \S\ref{sphysical}, \S\ref{spoles} with the result of
\cite{1311.1257} that the S-matrices of massless / BPS / special states are independent
of the choice of local coordinate system, to argue that to all orders in string perturbation 
theory the renormalized
physical masses are independent of the choice of local coordinate system and
that the residues at the poles associated with the unphysical / pure gauge sector 
states vanish. In other words
the poles in the S--matrix elements of massless / BPS / special states in the $-k^2$
plane occur only at the renormalized physical mass$^2$ defined in \S\ref{sphysical}.

The proof that physical masses are independent of the choice of local coordinates requires us
to assume that the corresponding physical states appear in the intermediate channel of the
S-matrix of some set of massless / BPS / special states. In the examples we have examined this
always seems to hold.

Finally in \S\ref{shetsup} we  briefly discuss generalization of our analysis to
Neveu-Schwarz sector states in heterotic and superstring theories.

\sectiono{A field theory example} \label{stoy}

In this section we shall illustrate the problem of mixing between physical and unphysical states
in a gauge theory. We shall also provide an algorithm for extracting the renormalized 
physical mass in this theory. This algorithm will be generalized to string theory in 
\S\ref{sphysical}.

\subsection{The model} \label{smodel}

Consider a quantum field theory in $D+1$ dimensions containing an abelian gauge
field $A_\mu$ and a pair of complex scalars $\phi, \chi$, each carrying charge $q$
under the gauge field. We consider a gauge invariant Lagrangian density of the form
\ben \label{eqft1}
\LL &=& -{1\over 4} F_{\mu\nu} F^{\mu\nu} - (\p_\mu\phi^* + i q A_\mu\phi^*) (\p^\mu\phi
-i q A^\mu \phi) - c \, (\phi^*\phi - v^2)^2  \nonumber \\
&& - (\p_\mu\chi^* + i q A_\mu\chi^*) (\p^\mu\chi
-i q A^\mu \chi) - V(\phi, \chi)\, , \nonumber \\
F_{\mu\nu} &\equiv & \p_\mu A_\nu - \p_\nu A_\mu\, ,
\een
where $V(\phi,\chi)$ is a potential 
whose detailed properties will be discussed shortly, but
for now we just mention that it plays no role in the breaking of the U(1) gauge symmetry.
Minimizing the potential in the first line we see that $|\phi|=v$ is the minimum of the
potential. We choose $\phi=v$ as the vacuum expectation value of $\phi$. 
We now define $\phi_{R,I},\chi_{R,I}$ via
\be
\phi= v + {1\over \sqrt 2} (\phi_R + i \phi_I), \quad \chi= {1\over \sqrt 2} (\chi_R + i \chi_I),
\ee
and
\be \label{edefm}
m\equiv \sqrt 2 \, q\, v\, .
\ee
We now describe the choice of the potential $V(\phi,\chi)$. We require it to have the
property that 
when expanded around the point $(\phi=v,\chi=0)$, it has an expansion of the form
\be \label{eqft2}
-{1\over 2} m_0^2 \chi_R^2 - {1\over 2} m^2 \chi_I^2 + \hbox{cubic and higher order terms
in $\phi_R,\phi_I,\chi_R,\chi_I$}\, ,
\ee
where $m_0$ is an arbitrary mass parameter but $m$ has been chosen to be the same 
quantity defined in \refb{edefm}.
Using this
we get, after throwing away total derivative terms,
\ben
\LL &=& -{1\over 2} \p_\mu A_\nu \p^\mu A^\nu -{1\over 2} m^2 A_\mu A^\mu + 
{1\over 2} (\p_\mu A_\mu -m\, \phi_I)^2 -{1\over 2} \p_\mu \phi_I \p^\mu \phi^I -{1\over 2} m^2 \phi_I^2
-{1\over 2} \p_\mu \phi_R \p^\mu \phi_R \nonumber \\
&& - 2\, c \, v^2 \, \phi_R^2
-{1\over 2} \p_\mu \chi_I \p^\mu \chi^I -{1\over 2} m^2 \chi_I^2
-{1\over 2} \p_\mu \chi_R \p^\mu \chi^R - {1\over 2}\, m_0^2 \, \chi_R^2 + \hbox{interaction terms}\, .
\nonumber \\
\een
To this we add a gauge fixing term
\be \label{egfixing}
\LL_{gf}=- {1\over 2} (\p^\mu A_\mu -m\, \phi_I)^2\, ,
\ee
so that the third term in $\LL$ is cancelled by $\LL_{gf}$ in the total Lagrangian density
$\LL+\LL_{gf}$. The resulting Lagrangian has the fields $A_\mu$, $\phi_I$ and $\chi_I$ all
carrying mass $m$, whereas $\phi_R$ and $\chi_R$ carry different masses.

Now if we work in the momentum space and are at the rest frame $k=(k^0,\vec 0)$
then the fields $A_i$ transform in the vector representation of the little group SO(D) 
whereas the fields $A_0, \phi_I$ and $\chi_I$ transform in the scalar representation of the
same group. At tree level the fields $A_i$ and $\chi_I$ are physical whereas the fields 
$A_0$ and $\phi_I$ are unphysical.\footnote{In the language that we shall develop shortly,
one linear combination of these fields will be called unphysical and the other will be called
pure gauge.}
In particular by choosing unitary gauge
we can remove $A_0$ and $\phi_I$ from the spectrum.
Alternatively
by choosing another gauge fixing term {\it e.g.}
$- (\p^\mu A_\mu -m\, \xi\, \phi_I)^2 / (2\xi)$ with $\xi\ne 1$ 
we could make the unphysical fields $A_0$ and $\phi_I$ have mass different from 
$m$ and hence non-degenerate with the physical fields. We shall however work with
$\xi=1$ and address the problems associated with the degeneracy directly since this is
what we shall need to do in string theory.
Our main goal will be to disentangle the physical and unphysical states after inclusion of
loop corrections. 

Now it is clear that under loop corrections the SO(D) vector fields $A_i$ cannot
mix with the unphysical fields and hence they remain physical states. These are the
analogs of the special states considered in \cite{1311.1257}. However the state $\chi_I$ can now mix
with $A_0$ and $\phi_I$. To see what kind of mixing is possible, we note that according the
general principle of gauge theory the corrections must take the form of a gauge invariant
term written in terms of the original variables $\phi$, $\chi$, $A_\mu$ together with a
possible renormalizaton of the gauge fixing term. Let us suppose that quantum corrections
generate a gauge invariant mass term for $\chi$ of the form $- \alpha\chi^* \chi$ and changes the
gauge fixing term \refb{egfixing} 
to $-(\p^\mu A_\mu -m\, \phi_I + \beta \phi_I + \gamma\chi_I)^2/2$.\footnote{We
could have also changed the coefficient of the $\p_\mu A^\mu$  inside the gauge
fixing term and added other gauge invariant terms, 
but the corrections we have taken are sufficiently general to
illustrate the basic points.}. Here, $\alpha$, $\beta$ and $\gamma$ are in principle computable constants which arise from loop corrections.
Adding these to \refb{eqft1} we can express the 
quadratic terms involving $A_\mu,\phi_I$ and $\chi_I$ as
\ben
&& -{1\over 2} \p_\mu A_\nu \p^\mu A^\nu -{1\over 2} m^2 A_\mu A^\mu 
 -{1\over 2} \p_\mu \phi_I \p^\mu \phi^I -{1\over 2} m^2 \phi_I^2
-{1\over 2} \p_\mu \chi_I \p^\mu \chi^I -{1\over 2} m^2 \chi_I^2
\nonumber \\
&& -{1\over 2} \, \alpha \, \chi_I^2 - \beta \phi_I \p_\mu A^\mu +{1\over 2} (2 m\beta-\beta^2) \phi_I^2
-\gamma \chi_I \p_\mu A^\mu -{1\over 2} \gamma^2 \chi_I^2 + (m-\beta) \, \gamma \, \phi_I \, \chi_I. 
\een
In momentum space, up to overall multiplication and momentum conserving delta functions,
the quadratic Lagrangian density in the $\vec k=0$ sector can be written as
\be \label{e1.8}
{1\over 2} A_i(-k) \{(k^0)^2 - m^2\} A_i(k) 
+ {1\over 2} 
\pmatrix{A_0(-k) & \phi_I(-k) & \chi_I(-k)} M \pmatrix{A_0(k) \cr \phi_I(k) \cr \chi_I(k)}\, ,
\ee
where
\be \label{emdef}
M=\pmatrix{ - (E^2 - m^2) &  {\bf i}\, E\, \beta & {\bf i} \, E\, \gamma    \cr  
-{\bf i}\, E\, \beta &   E^2 - (m-\beta)^2  &  (m-\beta)\gamma \cr 
- {\bf i} \, E\, \gamma  &  (m-\beta)\gamma  &  E^2 - m^2 - \gamma^2 -\alpha
}, \quad E\equiv k^0\, .
\ee
As expected $A_i(k)$'s, being special states, do not mix with other fields. In this example its mass
is not affected by the quantum corrections, but this is just a consequence of the limited number
of terms we have added, {\it e.g.} this could change if we had added a gauge invariant term
proportional to $F_{\mu\nu} F^{\mu\nu}$ in the quantum corrections to the Lagrangian density.

Let us define the matrices
\be \label{ext1}
\II = \pmatrix{-1& 0 & 0\cr 0 & 1 & 0\cr 0 & 0 & 1}, \quad
\wt F_T = \pmatrix{ 0 &  {\bf i}\, E\, \beta & {\bf i} \, E\, \gamma    \cr  
-{\bf i}\, E\, \beta &   2m\beta - \beta^2  &  (m-\beta)\gamma \cr 
- {\bf i} \, E\, \gamma  &  (m-\beta)\gamma  &  - \gamma^2 -\alpha
}, 
\ee
so that we can write
\be
M = - \{ (m^2 - E^2)\II - \wt F_T\}\, .
\ee
The full propagator (up to overall sign and factors of $i$) is then given by
\be \label{ext2}
\PP_T = - M^{-1} = \{ (m^2-E^2) \II - \wt F_T\}^{-1} \, ,
\ee
and the renormalized 
squared masses are the locations of the poles of this matrix in the $E^2$ plane.
Only one of these poles is physical. We need to find a systematic algorithm for determining
which one is physical and calculate its location.
This will be done in \S\ref{salgo}, but to facilitate the analysis we shall now introduce
a few notations.

Let us introduce a set of basis states as follows:
\be \label{ext3}
|p\rangle = \pmatrix{0\cr 0\cr 1}, \quad |g\rangle ={1\over |E|\sqrt 2} \pmatrix{-\bi E\cr |E|\cr 0}, 
\quad |u\rangle = {1\over |E|\sqrt 2} \pmatrix{\bi E\cr |E| \cr 0}\, .
\ee
The conjugate basis $\langle p|$, $\langle g|$ and $\langle u|$ are defined 
by taking transpose together
with a change of sign of the momentum vector. The latter operation changes the sign of
$E$ and hence effectively the conjugate basis corresponds to hermitian conjugates
of the vectors \refb{ext3}.
Then we have the following identities
\be \label{eidentity}
\pmatrix{\langle g|\II|g\rangle & \langle g|\II|u\rangle & \langle g|\II|p\rangle\cr
\langle u|\II|g\rangle & \langle u|\II|u\rangle & \langle u|\II|p\rangle\cr
\langle p|\II|g\rangle & \langle p|\II|u\rangle & \langle p|\II|p\rangle}
= \pmatrix{ 0 & 1 & 0\cr 1 & 0 & 0\cr 0 & 0 & 1}\, .
\ee
We shall call $|p\rangle$, $|g\rangle$ and $|u\rangle$ as tree level 
physical, pure gauge and 
unphysical states respectively. The name pure gauge for $|g\rangle$ stems from
the fact that on-shell (at $|E|=m$) 
this describes a pure gauge deformation of the vacuum at the linearized
level and the name physical originates from the fact that the $\chi_I$ field 
represented by the vector $|p\rangle$ is the
physical field at the tree level.

\subsection{The algorithm for computing the physical mass} \label{salgo}

Our goal will be to develop an algorithm for finding the corrected physical state
and the physical mass after taking into account the quantum correction to
$M$ represented by $\wt F_T$. Furthermore instead of aiming at the exact result we 
want to do this perturbatively in the parameters $\alpha,\beta,\gamma$ since this is
what we need in string theory.
The problem is made complicated by the fact that the full matrix $M$ is
expected to have zero eigenvalue at more than one value of $E$ near $m$, and we expect
only one of these to represent physical mass.
Let $m_p$ be the quantum corrected physical mass, and $|p\rangle'$ be the 
eigenvector with zero eigenvalue at
$E=m_p$. Then naively we might expect that as we switch off the perturbation
parameters $\alpha, \beta, \gamma$, the vector $|p\rangle'$ should approach the 
unperturbed physical state $|p\rangle$ and we can use this as a criterion for identifying the
quantum corrected physical state. The problem however is that since the unperturbed matrix
has three different eigenvectors with zero eigenvalue at $E=m$, what we have here is
an analog of degenerate perturbation theory and there is no guarantee that the eigenvectors
of the quantum corrected matrix will approach a particular unperturbed eigenvector in the
limit of switching off the perturbation. Indeed, we shall see that in general it is not possible
to construct an eigenvector with zero eigenvalue in the perturbed theory that
approaches the particular vector $|p\rangle$ in the limit $\alpha,\beta,\gamma\to 0$. 
The best we can do is to find such an
eigenvector that approaches a linear combination of the unperturbed physical state
$|p\rangle$ and the unperturbed pure gauge state $|g\rangle$ as we switch off the
perturbation. We shall take this as the criterion for identifying the quantum corrected
physical state and look for an algorithm for constructing such a state.

With this goal in mind, we now seek a change of basis of the form
\be \label{ech1}
|p\rangle' = A|p\rangle + B |g\rangle + C |u\rangle, \quad
|g\rangle' =  |g\rangle + D |p\rangle, \quad |u\rangle'=|u\rangle + K |p\rangle\, ,
\ee
such that the following conditions hold
\be \label{econd1}
\pmatrix{'\langle g|\II|g\rangle' & '\langle g|\II|u\rangle' & '\langle g|\II|p\rangle'\cr
'\langle u|\II|g\rangle' & '\langle u|\II|u\rangle' & '\langle u|\II|p\rangle'\cr
'\langle p|\II|g\rangle' & '\langle p|\II|u\rangle' & '\langle p|\II|p\rangle'}
= \pmatrix{ * & * & 0\cr * & * & 0\cr 0 & 0 & 1}\, ,
\ee
and
\be \label{econd2}
\pmatrix{'\langle g|\wt F_T|g\rangle' & '\langle g|\wt F_T|u\rangle' & '\langle g|\wt F_T|p\rangle'\cr
'\langle u|\wt F_T|g\rangle' & '\langle u|\wt F_T|u\rangle' & '\langle u|\wt F_T|p\rangle'\cr
'\langle p|\wt F_T|g\rangle' & '\langle p|\wt F_T|u\rangle' & '\langle p|\wt F_T|p\rangle'}
= \pmatrix{ * & * & 0\cr * & * & 0\cr 0 & 0 & *}\, .
\ee
where $*$ denotes unconstrained numbers.
Notice that \refb{ech1} is not the most general change of basis. In fact the most general
change of basis is related to the one given in \refb{ech1} by arbitrary mixing between the
states $|u\rangle'$ and $|g\rangle'$ without involving $|p\rangle'$. However all the
conditions demanded in \refb{econd1}, \refb{econd2} are invariant under such a change
of basis and hence by taking convenient linear combinations of $|u\rangle'$ and $|g\rangle'$
satisfying \refb{econd1}, \refb{econd2}
we can always ensure that the change of basis is of the form given in 
\refb{ech1}.
We now substitute \refb{ech1} into
\refb{econd1}, \refb{econd2} and use \refb{eidentity} to get
\ben \label{emaster}
&& A^* A+ B^* C + C^*B=1, \quad D^*A + C = 0, \quad K^*A+B=0\, , \nonumber \\
&& 
A\langle u|\wt F_T|p\rangle + B \langle u|\wt F_T|g\rangle + C \langle u|\wt F_T|u\rangle
+ K^* A \langle p|\wt F_T|p\rangle  + K^* B \langle p|\wt F_T|g\rangle + K^* C \langle p|\wt F_T|u\rangle
= 0 \nonumber \\
&& A\langle g|\wt F_T|p\rangle + B \langle g|\wt F_T|g\rangle + C \langle g|\wt F_T|u\rangle
+ D^* A \langle p|\wt F_T|p\rangle  + D^* B \langle p|\wt F_T|g\rangle + D^* C \langle p|\wt F_T|u\rangle
= 0\, . \nonumber \\
\een

We shall soon discuss how to construct $A,B,C,D,K$ perturbatively satisfying
\refb{emaster} and the criteria mentioned at the beginning of this subsection. However 
let us first examine the consequences of \refb{econd1} and \refb{econd2}. 
Using these equations we see
that in the primed basis the matrices $\II$ and $\wt F_T$ are exactly block diagonal, with the
$|p\rangle'$ block having no mixing with the $|u\rangle'$ and $|g\rangle'$ 
blocks. Of course the basis we have
chosen is not orthonormal in the $(|u\rangle',|g\rangle')$ sector, but this can be rectified by appropriate
linear transformation in the ($|u\rangle'$, $|g\rangle'$) space without affecting the 
$|p\rangle'$-$|p\rangle'$ element. Thus we get
\be \label{e2.19}
'\langle p|\PP_T|p\rangle' = 
\{ (m^2-E^2) - \wt F(E) \}^{-1} \, , 
\ee
where
\ben \label{edeffe}
\wt F(E) &\equiv& \, '\langle p|\wt F_T|p\rangle' 
\nonumber \\ &=& 
A^*A \langle p|\wt F_T|p\rangle + A^*B \langle p|\wt F_T|g\rangle + A^*C \langle p|\wt F_T|u\rangle +
B^*A \langle g|\wt F_T|p\rangle 
+ B^*B \langle g|\wt F_T|g\rangle \nonumber \\ &&
 + B^*C \langle g|\wt F_T|u\rangle 
+ C^*A \langle u|\wt F_T|p\rangle + C^*B \langle u|\wt F_T|g\rangle + C^*C \langle u|\wt F_T|u\rangle
\, . 
\een
The pole of \refb{e2.19} can be constructed iteratively by expressing this equation as
\be \label{eiterate}
E^2 = m^2 - \wt F(E),
\ee
and solving the equation iteratively by
starting with $E^2=m^2$. We can identify this as the physical pole provided the
following two conditions hold:
\begin{enumerate}
\item
Let us introduce a perturbation parameter $\lambda$ and take 
\be
\alpha \sim \lambda, \quad \beta\sim \lambda, \quad \gamma\sim\lambda\, .
\ee
In particular if $\alpha$, $\beta$, $\gamma$ arise at one loop order then the power of
$\lambda$ counts the number of loops.
We need to ensure  that the coefficient of $\lambda^n$ 
in the expressions for $A,\cdots K$ and $\wt F(E)$ are 
free from any pole at $E\simeq m$ for
every $n$. Otherwise the iterative procedure for finding the solution that starts
with $E=m$ will break down. 
\item We also need to ensure that the coefficient $C$ approaches 0 in the limit
$\lambda\to 0$ and $E\to m$ so that the state $|p\rangle'$ approaches a linear
combination of the tree level physical state and tree level pure gauge 
state in this limit.
$|p\rangle'$ will then satisfy the criteria mentioned at the beginning of this
subsection.
\end{enumerate}

We shall now discuss how to solve \refb{emaster} satisfying these conditions.
Since each matrix element of $\wt F_T$ is of order $\lambda$, we can factor
out the overall factor of $\lambda$ from the last two equations in \refb{emaster}, 
take the $\lambda\to 0$ limit, and regard \refb{emaster} as a set of $\lambda$ independent
equations which can be solved to determine the leading order result for
the coefficients $A,\cdots K$. It is easy to
check that leaving aside an overall phase there are as many unknowns as the number of
equations, and hence we expect these equations to have solutions.
Solving the leading order equations can in fact be facilitated by using
another expansion parameter, namely $(E^2-m^2)$. 
For this we note that  \refb{ext1}, \refb{ext3} gives
\ben \label{eorder}
&& \lambda^{-1} \langle p |\wt F_T|g\rangle \sim \OO(E^2-m^2) +\OO(\lambda), 
\quad \lambda^{-1} \langle g |\wt F_T|p\rangle\sim \OO(E^2-m^2) +\OO(\lambda), 
\nonumber \\ &&
 \lambda^{-1}\langle g |\wt F_T| g\rangle \sim \OO(E^2-m^2) +\OO(\lambda)\, ,
\een
while the other matrix elements of $\lambda^{-1}\wt F_T$ are of order unity as $E\to m$ 
and $\lambda\to 0$.
Making use of \refb{eorder}, let us
look for a leading order in $\lambda$
solution in which
\be \label{eabcdsim}
A, B, K \sim 1, \quad C, D\sim (E^2-m^2)\, .
\ee
Using \refb{eorder},  \refb{eabcdsim} we see that
to the leading order in $\lambda$, \refb{emaster} gives 
\ben \label{eleading}
&& A^*A=1 + \OO(E^2-m^2),  \quad \quad D^*A + C = 0, \quad K^*A+B=0, \nonumber \\ &&
\lambda^{-1}\left\{A\langle u|\wt F_T|p\rangle + B \langle u|\wt F_T|g\rangle 
+ K^* A \langle p|\wt F_T|p\rangle
\right\}=\OO(E^2-m^2)\, ,
\nonumber \\ &&
 \lambda^{-1}\left\{
 A\langle g|\wt F_T|p\rangle + B \langle g|\wt F_T|g\rangle + C \langle g|\wt F_T|u\rangle
+ D^* A \langle p|\wt F_T|p\rangle \right\} =\OO((E^2-m^2)^2)\, . \nonumber \\
\een
Each term in the left hand side of the
first, third and fourth equations is of order unity and each term in the left hand side of the
third and fifth equations is of order
$(E^2-m^2)$. The solution is
\ben \label{ecoeff}
&& A=1 + \OO(E^2-m^2), \quad 
K^* = \{ \langle u|\wt F_T|g\rangle - \langle p|\wt F_T|p\rangle\}^{-1}
\langle u|\wt F_T|p\rangle + \OO(E^2-m^2), 
\nonumber \\ && 
D^* = \{ \langle g|\wt F_T|u\rangle - \langle p|\wt F_T|p\rangle\}^{-1}
\{ \langle g|\wt F_T|p\rangle -K^* \langle g|\wt F_T|g\rangle\} +  \OO((E^2-m^2)^2)\, ,
\nonumber \\ &&
B = - K^*+ \OO(E^2-m^2), \quad C = - D^*+ \OO((E^2-m^2)^2)\, .
\een
Using \refb{eorder} and the comments below it, we see that as
long as the order $\lambda$ contribution to
$\{ \langle u|\wt F_T|g\rangle - \langle p|\wt F_T|p\rangle\}$ does not vanish
(and in particular does not have zero at $E^2=m^2$),
$A$, $B$ and $K$ given in \refb{ecoeff}
are of order unity, while $C$ and $D$ are of order $(E^2-m^2)$,
in agreement with our assumption \refb{eabcdsim}.
The reader may be surprised by the appearance of the one loop term
$\{ \langle u|\wt F_T|g\rangle - \langle p|\wt F_T|p\rangle\}$
in the denominator in a perturbation theory, but this is simply a consequence of the
degenerate perturbation theory that we need to carry out in this case. Requiring 
$\{ \langle u|\wt F_T|g\rangle - \langle p|\wt F_T|p\rangle\}$ to be non-zero  is
equivalent to demanding that the degeneracy between the physical and the 
unphysical / pure gauge states is lifted at the first order.
Starting with \refb{ecoeff} we can now iteratively solve the system of equations in a power 
series in $\lambda$ and $(E^2-m^2)$. 
For this we choose $A$ to be real,\footnote{Eqs.\refb{emaster} have a symmetry under which
the constants $A,B,C$ are multiplied by an overall phase. We have chosen this phase
appropriately to make $A$ real.
}
express eqs.\refb{emaster} as
\ben \label{emasteriteration}
&& A  = \sqrt{1 -B^*C-C^*B} \, , \nonumber \\
&& K^* = \{ \langle u|\wt F_T|g\rangle - \langle p|\wt F_T|p\rangle\}^{-1}
\bigg[A\langle u|\wt F_T|p\rangle + (B+K^*) \langle u|\wt F_T|g\rangle + C \langle u|\wt F_T|u\rangle
\nonumber \\ &&
+ K^* (A-1) \langle p|\wt F_T|p\rangle  + K^* B \langle p|\wt F_T|g\rangle + K^* C \langle p|\wt F_T|u\rangle
\bigg]\, ,
\nonumber \\
&& D^* = \{ \langle g|\wt F_T|u\rangle - \langle p|\wt F_T|p\rangle\}^{-1} \bigg[
A\langle g|\wt F_T|p\rangle + B \langle g|\wt F_T|g\rangle + (C+D^*) \langle g|\wt F_T|u\rangle
\nonumber \\ &&
+ D^* (A-1) \langle p|\wt F_T|p\rangle  + D^* B \langle p|\wt F_T|g\rangle + D^* C \langle p|\wt F_T|u\rangle
\bigg] \, ,
\nonumber \\
&& C = - D^*A, \quad B = - K^* A\, ,
\een
and evaluate the right hand sides of these equations iteratively, beginning with the leading
order solution. To get a perturbation expansion we also need to expand 
$ \{ \langle u|\wt F_T|g\rangle - \langle p|\wt F_T|p\rangle\}^{-1}$ in a power series in $\lambda$
starting with the leading order solution.
Each power of $\lambda$ will be free from any pole near $E^2=m^2$ as long as the
leading order result for $ \lambda^{-1}
\{ \langle u|\wt F_T|g\rangle - \langle p|\wt F_T|p\rangle\}$ does
not have any zero near $E^2=m^2$.
Once we determine the coefficients $A,\cdots K$ we can also determine $\wt F(E)$
using \refb{edeffe}.

Note that in this scheme even in a fixed order in $\lambda$ we need to iterate the
procedure infinite number of times to generate all powers of $E^2-m^2$. However eventually
we are interested in computing these coefficients at the physical mass$^2$ which differs
from $m^2$ by order $\lambda$. Similarly when we solve \refb{eiterate} to find the location
of the pole, we need to know the expansion of $\wt F(E)$ to order $(E^2-m^2)^n$ for computing the
correction to mass$^2$ to order $\lambda^{n+1}$. Thus for computing physical quantities
to any given order in $\lambda$ we need to run the iteration only a finite number of times.

We now observe that since eq.\refb{ecoeff} gives 
$B\simeq -K^* \sim 1$, it follows from \refb{ech1} that
$|p\rangle'$ differs from $|p\rangle$ by an order one term proportional to the 
pure gauge states. 
This is a consequence of having degenerate eigenvalues at the
tree level and will continue to be true in string theory as well.
On the other hand since $C\sim E^2-m^2$ which is of order $\lambda$ when
$E$ is set equal to the corrected physical mass,  the coefficient of $|u\rangle$ in
$|p\rangle'$ vanishes as $\lambda\to 0$. Thus the quantum corrected physical state 
approaches a linear combination of the unperturbed physical state and the
unperturbed pure gauge state in the limit in which we switch off the perturbation. This
is consistent with the criteria for identifying the quantum corrected physical state that we
set out at the beginning of this subsection.

\subsection{Explicit evaluation of the physical mass} \label{sexplicit}

Let us now explicitly evaluate the coefficients $A,\cdots K$ and $F(E)$ for the problem at hand
and from this find the location of the physical pole.
From \refb{ext1}, \refb{ext3}  it follows that here
\ben \label{efollows}
&& \langle p|\wt F_T|p\rangle = -\gamma^2-\alpha, \quad 
\langle g|\wt F_T|g\rangle = {\beta \over 2} ( 2m - \beta - 2|E|), \quad 
\langle u|\wt F_T|u\rangle = {\beta \over 2} ( 2m - \beta  + 2|E|), \nonumber \\ &&
\langle p|\wt F_T|g\rangle 
= \langle g|\wt F_T|p\rangle  ={1\over \sqrt 2} (-|E| + m-\beta)\gamma, \quad 
\langle p|\wt F_T|u\rangle
= \langle u|\wt F_T|p\rangle
= {1\over \sqrt 2} (|E| + m-\beta)\gamma,  \nonumber \\ &&
\langle g|\wt F_T|u\rangle
= \langle u|\wt F_T|g\rangle = {\beta  \over 2} ( 2m - \beta),
\een
This gives the leading order solutions \refb{ecoeff} to be
\ben \label{epert0}
&& A=1, \quad B = -K = -{1\over \sqrt 2} \, (\beta m  +\alpha)^{-1} \gamma (m+|E|), \nonumber \\ &&
C = -D = -(\beta m  +\alpha)^{-1}\left\{ {1\over \sqrt 2} (-|E| + m)\gamma
- {\beta \gamma\over \sqrt 2} ( m^2  - E^2) (\beta m  +\alpha)^{-1} \right\}\, .
\nonumber \\
\een
There are corrections to these solutions of order $\lambda$ and also of order $(E^2-m^2)$
($(E^2-m^2)^2$ in $C$ and $D$), but
these will not be needed for computing the leading correction to the physical mass.
Since $\alpha, \beta,\gamma$ are each
of order $\lambda$ we see that $B\simeq -K^*\sim 1$ and $C\simeq -D^* \sim (|E|-m)$ 
in the
$\lambda\to 0$ limit, in agreement with the general results quoted earlier.
Substituting these into \refb{edeffe} and using \refb{efollows} we get
\be 
\wt F(E) = -\alpha + \OO(\lambda^2) + \OO(\lambda) (|E|-m)\, .
\ee
The iterative procedure \refb{eiterate} now gives the leading order correction
to the physical mass
\be \label{epert1}
E^2 = m^2 + \alpha + \OO(\lambda^2)\, .
\ee
The physical state at leading order in $\lambda$, obtained from \refb{ext3}, \refb{ech1},
\refb{eleading}
and \refb{epert0} is given by, for $E=\sqrt{m^2+\alpha}$
\be \label{epert2}
\pmatrix{{i \gamma E}/(\alpha +\beta  m)\cr
- {\gamma  m}/(\alpha
   +\beta  m) \cr 1} + \OO(\lambda)\, .
\ee

Let us compare this with the exact result. We have from \refb{emdef}
\be \label{edetun}
\det M = - (E^2 - m^2 + m\beta)^2 (E^2 - m^2 -\alpha)\, .
\ee
This has zeroes at $E^2=m^2+\alpha$ and $E^2=m^2-m\beta$. Since we know that
$\beta$ enters through the renormalized gauge fixing term, the physical mass
should not depend on $\beta$. This determines the physical pole to be at
\be\label{exatc1}
E^2=m^2+\alpha\, ,
\ee 
which agrees with the perturbative result \refb{epert1}.
Furthermore at $E=\sqrt{m^2+\alpha}$ we can easily compute the zero eigenvector of
$M$ and it is given by
\be \label{exact2}
v = \pmatrix{ {i \gamma  E}/(\alpha +\beta  m)\cr
- {\gamma  m}/(\alpha
   +\beta  m) \cr 1}\, .
\ee
This agrees with the perturbative result \refb{epert2} up to corrections of order $\lambda$.

\subsection{Masses of the unphysical / pure gauge states} \label{sunp2}

For completeness we shall also describe the computation of the masses in the 
unphysical / pure gauge sector using perturbation theory.  For this we define the matrices
\be
\II' = \pmatrix{ ~'\langle g|\II |g\rangle' & ~'\langle g|\II |u\rangle'\cr
~'\langle u|\II |g\rangle' & ~'\langle u|\II |u\rangle'}\, , \quad
\wt F' = \pmatrix{ ~'\langle g|\wt F_T |g\rangle' & ~'\langle g|\wt F_T|u\rangle'\cr
~'\langle u|\wt F_T |g\rangle' & ~'\langle u|\wt F_T|u\rangle'}\, .
\ee
Then the unphysical / pure gauge sector masses will be at the zeroes of the 
eigenvalues of the
matrix
\be \label{eunp3}
(m^2 - E^2)\II' - \wt F'(E)\, ,
\ee
as a function of $E$.\footnote{It follows from \refb{ech1}, \refb{eabcdsim} that $\II'$ is a
non-singular matrix near $E\sim m$ and hence the zero eigenvalue of \refb{eunp3} occurs
at the same value of $E$ as that of $m^2 - E^2 - (\II')^{-1} \wt F'(E)$.}
For computing the first subleading correction to the unphysical mass we can 
use the ansatz
that the zero eigenvalue of \refb{eunp3} will occur at $(E-m)\sim\lambda$ and evaluate
each matrix element to order $\lambda$ using this ansatz.  
Since $(m^2-E^2)\sim\lambda$
we have to evaluate $\II'$ to order unity.  It follows from \refb{ech1} and the fact that at the
leading order $D\sim (E^2-m^2)\sim \lambda$ that $\II'$ to order $\lambda^0$ has the structure
\be 
\pmatrix{0 & 1 \cr 1 & K^* K}\, ,
\ee
with $K$ given in \refb{ecoeff}.
On the other hand \refb{ech1}, \refb{eorder} and \refb{eabcdsim} shows that $\wt F'$ to
order $\lambda$ has
the structure
\be 
\pmatrix{ 0 &  \langle g|\wt F_T|u\rangle\cr
\langle u|\wt F_T |g\rangle & ~'\langle u|\wt F_T|u\rangle'}\, .
\ee
Thus for computing order $\lambda$ correction to the unphysical / pure gauge sector
masses we need to look for zero eigenvalue of the matrix
\be\label{eunp4}
\pmatrix{ 0 & (m^2 - E^2) - \langle g|\wt F_T |u\rangle\cr
(m^2 - E^2) - \langle u|\wt F_T |g\rangle  & (m^2-E^2) K^*K - ~'\langle u|\wt F_T|u\rangle'}\, .
\ee
Now in order that a matrix has zero eigenvalue, its determinant must vanish. From
the structure of the matrix given above it is clear that this requires one of
the off-diagonal elements
to vanish. Since the off-diagonal elements are conjugates of each other and hence
vanish at the same value of $E$, the condition for zero eigenvalue of the \refb{eunp4} can 
be stated as
\be
(m^2 - E^2) - \langle u|\wt F_T |g\rangle = 0\, .
\ee
Using the value of $\langle u|\wt F_T |g\rangle$ quoted in \refb{efollows} we
see that to order $\lambda$ the renormalized masses in the unphysical / pure gauge 
sector 
occur at the zero of
\be 
E^2 - m^2 + m\beta = 0\, .
\ee
This is in agreement with the exact result quoted below \refb{edetun}.

\sectiono{Organization of off-shell amplitudes in string theory} \label{sorganize}

In this section we shall discuss some general aspects of off-shell states and
off-shell amplitudes in closed bosonic string theory.

\subsection{Off-shell string states and a basis} \label{sbasis}

We begin by describing the space of off-shell string states with which we shall work and
reviewing some well known results about the choice of basis for off-shell states.
Off-shell string states are required to satisfy the following conditions:
\begin{enumerate}
\item They have ghost number 2 where we count the $c$, $\bar c$ ghosts to have
ghost number 1, $b$, $\bar b$ ghosts to have ghost number $-1$ and SL(2,C) invariant
vacuum to have ghost number 0.
\item They are annihilated by the $b$, $\bar b$ ghost zero modes $b_0$ and $\bar b_0$ and 
$L_0-\bar L_0$ where $\bar L_n$ and $L_n$ are the total 
left and right moving Virasoro generators.
\end{enumerate}
This is also the space of off-shell states in covariant closed string field theory in the
Siegel gauge\cite{9206084}. The requirement of annihilation by $(L_0-\bar L_0)$ and
$(b_0-\bar b_0)$ is needed for consistently defining off-shell amplitude\cite{Nelson}
whereas the condition $(b_0+\bar b_0)\, |\hbox{state}\rangle=0$ is needed to make
the kinetic operator invertible.\footnote{In contrast the off-shell states in gauge invariant closed
string field theory of \cite{9206084} are only annihilated by $(b_0-\bar b_0)$
and $(L_0-\bar L_0)$. Like in all gauge theories, the kinetic operator in this
theory is not invertible till we fix a gauge and the Siegel gauge condition
of annihilation by $(b_0+\bar b_0)$ precisely does that. In quantum closed string field theory we
also need to relax the constraint on the ghost number and allow states of all ghost numbers to 
propagate in the loop. In our analysis we shall dump all the loop contributions into one particle
irreducible (1PI) amplitudes and express the full amplitude as sum of tree diagrams 
constructed out of 1PI amplitudes as vertices and 
tree level propagators. Thus the only place where we have to explicitly introduce
off-shell states is as the external lines of the 1PI amplitudes and as the states propagating along
the propagator in the tree amplitudes. These states always carry ghost number two when we 
compute physical amplitudes relevant for mass renormalization or S-matrix elements, and
hence we have put that restriction on the definition of off-shell states.
}
In this space we can introduce a non-degenerate inner product between states $|s\rangle$ and
$|s'\rangle$ via
\be \label{einnerp}
\langle s| s'\rangle \equiv \langle s| c_0\bar c_0 |s'\rangle_{BPZ}
\ee
where $\langle r|r'\rangle_{BPZ}$ is the BPZ inner product. 
In defining the bra $\langle r|$ corresponding to a given ket $|r\rangle$ we reverse the
sign of the momentum. We also remove the momentum conserving delta function from the
definition of the inner product.
The fact that the inner product is
non-degenerate follows from the Fock space representation of the basis states.

On-shell condition for the string state $|s\rangle$ takes the form
\be
L_0|s\rangle =0\, ,
\ee 
which also implies $\bar L_0|s\rangle=0$. On-shell we can divide the states into physical,
pure gauge and unphysical states as follows. First of all pure gauge states are of the form
\be 
Q_B |r\rangle
\ee
where $Q_B$ is the total BRST charge (left moving plus right moving) and
$|r\rangle$ is a state of ghost number 1  annihilated by $b_0$, $\bar b_0$,
$L_0$ and $\bar L_0$.
Since $Q_B$ has ghost number 1, commutes
with $L_n$, $\bar L_n$ and $\{Q_B, b_0\}=L_0$
and $\{Q_B, \bar b_0\} = \bar L_0$, it follows that
$Q_B|r\rangle$ has ghost number 2 and is  annihilated by $b_0$, $\bar b_0$,
$L_0$ and $\bar L_0$. 

Physical states are defined to be states of ghost number two 
which are annihilated by $Q_B$,
$b_0$, $\bar b_0$,
$L_0$ and $\bar L_0$
 but cannot
be written in the form $Q_B|r\rangle$ with $|r\rangle$ annihilated by $b_0$, $\bar b_0$,
$L_0$ and $\bar L_0$. It follows from this that the physical states are
orthogonal to pure gauge states. The main point to note is that 
$\{Q_B, c_0\}$ and
$\{Q_B, \bar c_0\}$ do not have any $c_0$ or $\bar c_0$ factor, and hence
the matrix elements of $\{Q_B, c_0\}$ and
$\{Q_B, \bar c_0\}$ between states,
satisfying condition 2 above, vanish. The same argument,
together with the relation $Q_B^2=0$, shows that
the pure gauge states also have vanishing inner product with pure gauge states.
A linearly independent basis of physical states is the maximal set of physical states
satisfying the
condition that no linear combination of these basis states is a pure gauge state.

Now since the inner product is non-degenerate there must exist states which have
non-vanishing inner product with the pure gauge states. 
These states are annihilated by $b_0$, $\bar b_0$, $L_0$ and $\bar L_0$, but not
by $Q_B$.
We shall call them 
unphysical states. 
We can choose a linearly 
independent
basis of unphysical states such that no linear combination is annihilated by $Q_B$. 
The number of such basis states must
be at least equal to the number of pure gauge states so that we have a non-degenerate 
inner product matrix.  We shall
now argue that the number is actually equal to the number of pure gauge states. For this
let us temporarily relax the constraint on the ghost number and consider states of all
ghost number annihilated by $b_0$, $\bar b_0$, $L_0$ and $\bar L_0$. Then since for
every unphysical state $|s\rangle$ of ghost number $g$,
$Q_B|s\rangle$ is a pure gauge state of ghost number $g+1$, we conclude that the
number of pure gauge states at ghost number $g+1$ is the same as the number of
unphysical states at ghost number $g$. On the other hand, since the inner product
\refb{einnerp} pairs states of ghost number $g$ and $4-g$,
 we know from our previous argument that the number of unphysical states at ghost
 number $3-g$ must be at least equal to the number of pure gauge states at ghost
 number $g+1$ and hence the number of unphysical states at ghost number $g$.
 Taking $g\to 3-g$ we can arrive at the reverse conclusion. This  shows that the number of
 unphysical states at ghost number $3-g$ should be equal to the number of unphysical states
 at ghost number $g$ and hence the number of pure gauge states at ghost number
 $g+1$. Taking $g=1$ we see that the number of unphysical states at ghost number 2
 must be equal to the number of pure gauge states at ghost number 2.
 This is the promised  result. 
 
 Let us now return to states of ghost number 2 only.  
 We have already seen that the inner product pairs unphysical states with pure gauge
 states by a non-degenerate matrix and that the pure gauge states are orthogonal
 to themselves as well as physical states. By adding appropriate linear combinations
 of pure gauge states and physical states to the unphysical states we can ensure that 
 the latter are orthonormal to the physical states and unphysical states. 
 Taking further linear combinations within physical states and within unphysical states
 we can ensure that the physical states form an orthonormal basis and that
 the pure gauge states
 and the unphysical states are paired in a one to one fashion.
 Thus at any mass level the inner
 product matrix will have a block diagonal structure of the form
 \be \label{einn2}
 \II = \pmatrix{ & I & \cr I & & \cr & & I}
 \ee
 where $I$ denotes identity matrix of appropriate dimensions. The first set of rows/columns
 stand for pure gauge states, the second set of rows/columns stand for unphysical states
 and the last set of rows/columns stand for physical states. At non-zero momentum,
 it is in fact possible to choose
 a basis satisfying this requirement with physical states of the form
 \be \label{ephysbasis}
 |\alpha\rangle = c_1 \bar c_1 |\Phi_\alpha\rangle
 \ee
 where $\Phi_\alpha$ are dimension (1,1) primary in the matter sector satisfying
 \be
 \langle \alpha| \beta\rangle \equiv
 \langle \Phi_\alpha |c_{-1}\bar c_{-1} c_0\bar c_0 c_1\bar c_1|\Phi_\beta\rangle_{BPZ} = \delta_{\alpha\beta}\, .
 \ee
 Physical states of the form \refb{ephysbasis} are dimension zero primaries and hence 
 transform as scalars under conformal transformation.

So far we have reviewed well known results, but now we shall make a small jump and
discuss the off-shell continuation of these results.
At a given mass level $m$ we can go off-shell (satisfying the two conditions mentioned
at the beginning of this section) by deforming the momentum $k$ such that
$k^2+m^2$ is deformed away from zero. We shall require the deformed basis to still
satisfy the inner product structure described in \refb{einn2}, but will need to relax the
various other requirements by terms of order $(k^2+m^2)$.
For example if we take a state $|s\rangle$ 
of ghost number 1 that is annihilated by $b_0$, $\bar b_0$ and $(L_0-\bar L_0)$,  
and apply the BRST charge $Q_B$ on it, the resulting state will not be
annihilated by $b_0$ and $\bar b_0$. The part that is not annihilated by $b_0$ and
$\bar b_0$ is given by $(c_0+\bar c_0) L_0 |s\rangle = {1\over 4}
(k^2+m^2) (c_0+\bar c_0) |s\rangle $. Hence the off-shell
`pure gauge' states
will have to be defined as $Q_B|s\rangle - {1\over 4}
(k^2+m^2) (c_0+\bar c_0) |s\rangle$. These are not annihilated by $Q_B$ but under the
action of $Q_B$ give states proportional to $(k^2+m^2)$. Similarly 
physical states will now be defined by first continuing the momentum off-shell and then
by adding appropriate linear combination of unphysical states proportional to
$(k^2+m^2)$ so that they remain orthonormal to the pure gauge states.
These
will only be BRST invariant up to terms of order $(k^2+m^2)$ and transform under
a conformal transformation as scalars up to terms of order $(k^2+m^2)$. Similar
procedure can be used to define the unphysical states off-shell so that they remain
orthogonal to physical states and themselves.

We shall denote by $|\alpha\rangle_p$, $|s\rangle_g$ and $|s\rangle_u$ an appropriate
basis of off-shell
physical, pure gauge and unphysical states at mass level $m$, satisfying the
identities
\be \label{eidentitystring}
 ~_p\langle\alpha|\beta\rangle_p=\delta_{\alpha\beta}, \quad 
~_g\langle r|s\rangle_u=~_u\langle r|s\rangle_g=\delta_{rs},
\quad ~_p\langle\alpha|s\rangle_u = ~_p\langle\alpha|s\rangle_g
= 0, \quad ~_g\langle r|s\rangle_g = ~_u\langle r|s\rangle_u=0\, .
\ee
Note that this preserves the inner product matrix $\II$ given in \refb{einn2}.
We shall see that at higher loop order we need to redefine the physical, unphysical and
pure gauge states by making a further rotation of the basis.

\subsection{Off-shell amplitudes} \label{soff}

In this subsection we shall describe the construction of off-shell amplitudes in string
theory following \cite{1311.1257}, which in turn was inspired by bosonic 
string field theory\cite{9206084} and other earlier work ({\it e.g.} \cite{Nelson,divecchia}).
In order to define off-shell amplitudes in string theory we need to introduce local coordinate
system around the punctures on the Riemann surface where the vertex operators are 
inserted\cite{Nelson} (see also \cite{Cohen:1985sm,Cohen:1986pv,AG1,AG2}).
Let us denote by $z$ a reference coordinate system on a Riemann surface, possibly consisting of
several coordinate charts. Let $z_i$ denote the location of the $i$-th puncture in the $z$-coordinate
system and $w_i$ denote the local coordinate system around the $i$-th puncture,
related to $z$ by some functional relation $z=f_i(w_i)$ such that the $w_i=0$ point gets mapped
to $z=z_i$: $f_i(0)=z_i$. Then the contribution to the $n$-point off-shell amplitude from the
genus $g$  Riemann surfaces can be expressed as
\be
\int_{\MM_{g;n}}  \left\langle \prod_{i=1}^n f_i\circ V_i(0) \, \times \, 
\hbox{ghost insertions}\right\rangle\, ,
\ee
where $f\circ V(0)$ denotes the conformal transformation of the vertex operator $V$ by the
function $f(w)$, the correlator $\langle ~\rangle$ is evaluated in the reference 
$z$-coordinate system and $\int_{\MM_{g;n}}$ denotes integration over the moduli space of
Riemann surfaces of genus $g$ with $n$ punctures with appropriate measure. 
A detailed description of how to construct the integration measure (or equivalently the rules for
inserting $b$-ghosts into the correlation function) for a given choice of local coordinate system
can be found in \cite{Nelson,9206084}.
The off-shell
amplitudes defined this way depend on the choice of local coordinate system $w_i$ but 
are independent
of the choice of the reference coordinate system $z$.

We shall work
with a class of local coordinate systems satisfying the following 
properties:\footnote{We note that the choice of local coordinates which appear in the
Siegel gauge amplitudes in closed bosonic string field theory of \cite{9206084}
automatically satisfies these requirements. Thus all our subsequent discussions hold for
this theory. In particular our analysis shows that the renormalized physical masses are
the same in different versions of closed string field theory using different vertices satisfying
Batalin-Vilkovisky equations. Since these different versions are related to each other by
field redefinitions together with a change in the gauge fixing condition\cite{9301097} this indirectly
tests gauge invariance of the renormalized physical masses in closed string field theory.}
\begin{enumerate}
\item \label{i1} 
The local coordinate system is taken to be symmetric in all the puncture, \i.e.\ the
function $f_i(w)$ should depend on $i$ only via the location $z_i$ of the puncture.
\item On 3-punctured sphere and 1-punctured tori the choice of the local coordinate system 
is arbitrary subject to condition \ref{i1}. We declare all 3-punctured spheres and 1-punctured
tori to be one particle irreducible (1PI) contributions to genus zero 3-point amplitudes and
genus one 1-point amplitudes respectively.
\item
We can construct a set of 4-punctured spheres
by gluing a 3-punctured sphere with another 3-punctured sphere at one each of their
punctures by the plumbing fixture procedure
\be \label{egluing}
w_1 w_2 = e^{-s+i\theta} \, \quad 0\le \theta < 2\pi, \quad 0\le s<\infty\, .
\ee
Here $w_1$ and $w_2$ are the local coordinates at the punctures used for gluing.
We choose the local coordinates  on these 4-punctured
spheres to be the ones induced from the local coordinates on the original 
3-punctured spheres\cite{peskin},
and
declare
the contribution from these 4-punctured spheres to off-shell four point amplitudes
to be the one particle reducible (1PR)
contributions to the genus zero four point amplitudes. On the rest of the genus zero four
punctured Riemann surfaces
we choose the local coordinate system arbitrarily subject to condition \ref{i1} and
continuity and declare them to be 1PI contributions 
to genus zero four point amplitude. We shall use a shorthand notation calling the corresponding
Riemann surfaces 1PI Riemann surfaces.
Similarly by gluing a 3-punctured sphere to a 1-punctured torus we can 
generate a set of 2-punctured tori. We choose the local coordinates
on these 2-punctured tori to be the ones induced from the local coordinates
of the 3-punctured sphere and the 1-punctured torus, and declare their contribution to be the
1PR contribution to the genus one 2-point function. On the rest of the
2-punctured tori we choose the local coordinates arbitrarily subject to condition \ref{i1} and
the requirement of continuity, and declare them to be 1PI contribution to the genus one 2-point
amplitude. 
\item We now repeat this process to Riemann surfaces of higher genus and/or higher number of
punctures. At any stage, Riemann surfaces which can be obtained by gluing two or more
1PI Riemann surfaces
to each other using the plumbing fixture procedure are declared to be  contributions to 1PR amplitudes
and on these Riemann surfaces the choice of local coordinates is induced from the local coordinates
of the 1PI Riemann surfaces which have been glued. The rest of the Riemann surfaces are
declared as 1PI contributions and the local coordinates at the punctures
on these Riemann surfaces
can be chosen arbitrarily subject to condition \ref{i1} and continuity.
\end{enumerate}
We shall call the choice of local coordinates satisfying the criteria described above 
{\it `gluing compatible local coordinate system'}.  In the language of string field theory this has
been called off-shell factorization, -- a brief discussion and relevant references can
be found in \cite{0708.2591}.

For our analysis it will also be useful to introduce the notion of amplitudes which are
{\it 1PI in a given momentum $k$}, where $k$ is the sum of a subset of the momenta carried by
the external states of that amplitude. 
Riemann surfaces 1PI in the leg carrying
momentum $k$ are defined to be those Riemann surfaces which cannot be obtained by gluing
two or more 1PI or 1PR 
Riemann surfaces at punctures carrying momenta $k$ and $-k$.
Thus this set of Riemann surfaces include the usual 1PI Riemann surfaces but also many 1PR
Riemann surfaces which are obtained by gluing two or more 1PI Riemann surfaces at
punctures carrying momenta other than $k$ or $-k$.
The total contribution to an amplitude 1PI in momentum $k$ is then obtained by integrating over
the moduli spaces of all Riemann surfaces which are 1PI in momentum $k$.

As an example consider genus one 2-point function with external vertex operators carrying
momentum $k$ and $-k$. This receives contribution from 1PI Riemann surfaces and also
1PR Riemann surfaces obtained by gluing 1-punctured torus to 3-punctured sphere. However all
of these are counted as 1PI in the momentum $k$ since the 1PR Riemann surfaces are 
obtained by gluing punctures carrying zero momentum, and not momentum $\pm k$.

\begin{figure}
\begin{center}
\figa
\end{center}
\vskip -1in
\caption{Pictorial representation of eq.\refb{eprop}. \label{f1}}
\end{figure}

\subsection{Off-shell amplitudes from 1PI amplitudes} \label{soff1PI}

As we shall now discuss, the off-shell amplitudes constructed with the help of such choice
of local coordinates can be organized in the same way that the full amplitudes in a quantum field
theory can be organized as sums over tree level Feynman diagrams with 1PI amplitudes as
vertices. As in \cite{1311.1257}  we begin our discussion with the propagator.
We shall 
work with general off-shell string states of ghost number 2, as defined in \S\ref{sbasis}.
If $\pm k$ denote the momenta carried by the external legs, then 
let $\wh\FF$ be the contribution to the off-shell two point amplitude from Riemann surfaces
which are 1PI in momentum $k$. This includes
sum over different genera starting from genus 1. As discussed in \cite{1311.1257}, this
can be regarded as a map from $\HH\times \HH$ to ${\mathbb C}$
where $\HH$ denotes 
the Hilbert space of off-shell states of
ghost number two as defined in \S\ref{sbasis},
but using the
duality between ghost number two and ghost number four states by the BPZ inner
product we can also regard this as a map from states of ghost number two to states of
ghost number four which are annihilated by $c_0$ and $\bar c_0$. 
We can include a further action by $\bar b_0 b_0$ to regard
$\wh \FF$ as a map from $\HH$ to $\HH$. 
This is the viewpoint we shall adopt from now.
The factor of $\bar b_0 b_0$ in fact arises naturally
in the tree level propagator of the string, which after being stripped of this factor, has
the form
\be \label{eDelta}
\Delta = {1\over 4\pi} \int_0^\infty ds \int_0^{2\pi} d\theta \, e^{-s (L_0+\bar L_0)} 
e^{i\theta (L_0-\bar L_0)} = {1\over 2 (L_0+\bar L_0)} \delta_{L_0,\bar L_0}\, .
\ee
With this convention the full propagator is given by
\be \label{eprop}
\Pi = \Delta + \Delta \wh \FF \Delta + \Delta \wh \FF \Delta \wh \FF \Delta  + \cdots
= \Delta (1 - \wh\FF\Delta)^{-1} = (1-\Delta \wh\FF)^{-1} \Delta\, .
\ee
Pictorially this contribution can be represented as in Fig.~\ref{f1} with 
the horizontal line denoting $\Delta$ and the blob marked 1PI denoting the
contribution $\wh\FF$ from the Riemann surfaces that are 1PI in momentum $k$. 

\begin{figure}
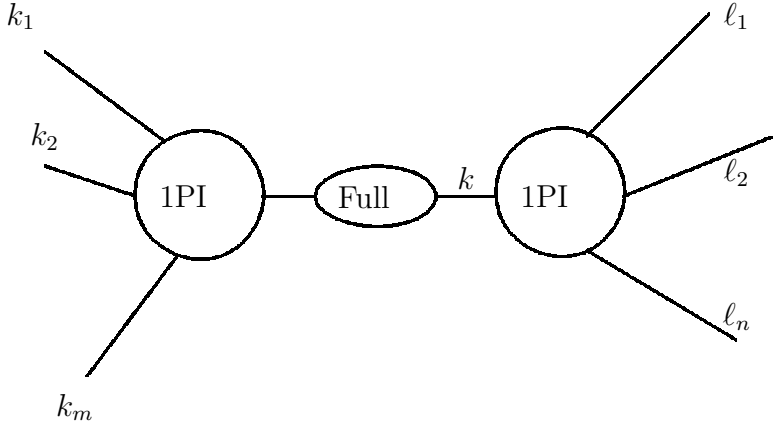

\begin{center}
\figb
\end{center}
\vskip -.5in
\caption{Pictorial representation of the second terms on the right hand sides of
eq.\refb{eGamma}. Here 1PI means sum of
contributions which are 1PI in the leg carrying momentum $k$, whereas  Full means sum
of all contributions to the 2-point function shown in Fig.\ref{f1}. \label{f2}}
\end{figure}

If $\FF$ is the full
off-shell two point function, then $\FF$ and $\Pi$ are related by
\be \label{estringprop}
\Pi = \Delta + \Delta \FF\Delta\, .
\ee
Also $\FF$ and $\wh\FF$ are related by
\be \label{edefff1}
\FF = \wh\FF + \wh\FF \Delta \wh\FF + \cdots =
\wh \FF (1 - \Delta \wh\FF)^{-1} = (1-\wh\FF\Delta)^{-1} \wh\FF =
\wh\FF + \wh\FF (\Delta^{-1} - \wh\FF)^{-1} \wh\FF\, .
\ee
Like $\wh\FF$, 
$\FF$, $\Pi$ and $\Delta$ can be regarded as maps from  $\HH$ to $\HH$.

As described in \cite{1311.1257}, we can use \refb{edefff1} to define $\wh\FF$ in terms
of $\FF$.
At genus one $\wh\FF=\FF$. Starting with this, we define $\wh\FF$ at genus two
so as to satisfy \refb{edefff1} up to genus two. Physically the contribution
to $\wh\FF$ at genus two is given by integrating over those Riemann surfaces
which cannot be obtained by plumbing fixture of a pair of genus one Riemann
surfaces. This definition of course depends on the choice of local coordinates
at the punctures that we use to glue the two genus one Riemann surfaces. This
procedure can be continued to define $\wh\FF$ at higher orders.

As another example let us consider an $m+n$ point amplitude 
$\Gamma$ 
with external momenta
$k_1,\cdots k_m$, $\ell_1,\cdots \ell_n$ satisfying $\sum_{i=1}^m k_i = - \sum_{j=1}^n \ell_j = k$,
and other quantum numbers $a_1,\cdots a_m$, $b_1,\cdots b_n$. 
Our goal is to express the amplitude in a way that makes manifest the poles in the
momentum $k$. For this we introduce two auxiliary quantities: $\Gamma_1^a$ describing the
contribution to $(m+1)$-point functions with external states carrying quantum numbers
$a_1,\cdots a_m$, $a$ and momenta $k_1,\cdots k_m$, $-k$ and
$\Gamma_2^b$ describing the
contribution to $(n+1)$-point functions with external states carrying quantum numbers
$b_1,\cdots b_n$, $b$ and momenta $\ell_1,\cdots \ell_n$, $k$. Here the quantum numbers
$a$ and $b$ run over all off-shell string states of ghost number 2.
Note that we have not explicitly exhibited the dependence of $\Gamma_1$ on the indices
$a_1,\cdots a_m$ and momenta $k_1,\cdots k_m$ for brevity; a similar comment holds for
$\Gamma_2$.
We shall also introduce the quantities $\wh\Gamma_1^a$ and $\wh\Gamma_2^b$ which 
describe contributions to $\Gamma_1^a$ and $\Gamma_2^b$ from those Riemann surfaces
which are 1PI in the leg carrying momentum $k$ (in the sense described at the end of
\S\ref{soff}).
Then the full contribution to $\Gamma$
can be expressed as
\ben \label{eGamma}
\Gamma &=& \wh\Gamma + \wh\Gamma_1^a \II_{ac} \Pi_{cb} \wh \Gamma_2^b 
\nonumber \\
&=& \wh\Gamma + 
\wh\Gamma_1^T \II 
\Delta (1 - \wh\FF\Delta)^{-1} \wh\Gamma_2 \nonumber \\
&=&  \wh\Gamma + \wh\Gamma_1^T \II 
(1-\Delta \wh\FF)^{-1} \Delta \wh\Gamma_2
 \, ,
\een
where $\wh\Gamma$ represents contributions to $\Gamma$ which are 1PI in the
leg carrying momentum $k$ and $\II$ is the inner product matrix \refb{einn2} over the full
space of off-shell string states. The equality between different expressions on the right hand
sides of \refb{eGamma} follows from \refb{eprop}.
A pictorial representation of the second term on the right hand side of the first line of
\refb{eGamma}
has been shown in Fig.~\ref{f2}

\sectiono{Physical state propagator in string theory} \label{sphysical}

In this section we shall generalize the gauge theory analysis of \S\ref{stoy} to give
an iterative procedure for constructing physical state propagator in string theory.
From this we can compute the masses of physical states.

\subsection{Renormalized propagator at a given mass level} \label{s4.1}

Since string theory contains infinite number of states, the quantities $\Pi$, $\Delta$,
$\FF$ and $\wh\FF$ introduced in \S\ref{soff1PI} are all infinite dimensional matrices.
Our first step will be to
`integrate out' all states except the ones at mass level $m$ so that we can
work with finite dimensional matrices with rows and columns labelled by states at
mass level $m$.\footnote{Throughout this
paper we shall denote by states at mass level $m$ all states which have tree level
mass $m$, \i.e.\ states which are annihilated by $L_0$ and $\bar L_0$ when
$k^2=-m^2$.}
For this
we denote by $P_T$ the total projection operator at mass level $m$,
\be \label{edefpt}
P_T = \{|\alpha\rangle_p ~_p\langle \alpha| + |s\rangle_g ~_u\langle s|
 + |s\rangle_u ~_g\langle s| \}  \, ,
\ee
and define 
\be \label{edefbardelta}
\bar\Delta = \Delta - (k^2+m^2)^{-1} P_T\, ,
\ee
%We also define $\bar\FF$ through
\be \label{edefbarff}
\bar\FF =  \wh\FF + \wh \FF \bar\Delta \wh \FF +\cdots 
=  \wh \FF (1 - \bar\Delta \wh\FF)^{-1} 
=   (1 - \wh\FF \bar\Delta)^{-1} \wh\FF %\nonumber \\
= \wh \FF + \wh\FF (\bar\Delta^{-1} - \wh\FF)^{-1} \wh\FF 
\, ,
\ee
where $\wh\FF$ has been defined in \S\ref{soff1PI}.
It is clear from the definition of $\bar\Delta$ and $\wh\FF$
that their genus expansions do not have any poles 
at $k^2=-m^2$. Hence $\bar\FF$ defined in \refb{edefbarff}
also does not have such poles. From \refb{edefff1}, \refb{edefbarff}
we get
\be \label{eweget}
\FF =\bar \FF \{ 1 - (k^2+m^2)^{-1} P_T \bar\FF\}^{-1} = \{ 1 - \bar \FF\, (k^2+m^2)^{-1} P_T\}^{-1}
\bar \FF\, .
\ee

We now define
\be \label{edefppt}
\PP_T =  P_T \, \II  \, \Pi \, P_T, \quad \wt F_T = P_T \, \II \, \bar \FF P_T, \quad
F_T = P_T \II  \FF P_T 
\, ,
\ee
where $\Pi$ has been defined in \refb{eprop}. Physically $F_T$ denotes the two point
amplitude restricted to external states of mass level $m$, $\wt F_T$ is the
contribution to $F_T$ that is 1PI in momentum $k$ 
{\it after integrating out all states other than those at mass level
$m$}, and $\PP_T$ 
denotes the off-shell two point Green's function 
restricted to external states of mass level $m$.
It  follows from \refb{eweget}, \refb{edefppt} that
\ben \label{edefpr}
F_T &=& \wt F_T (1 - (k^2+m^2)^{-1} \II \wt F_T)^{-1}, \nonumber \\
\PP_T &=& (k^2+m^2)^{-1} \II  P_T+ (k^2+m^2)^{-2} \, F_T = P_T \{ (k^2+m^2)\II - \wt F_T\}^{-1}  \, ,
\een
where it is understood that the inverse on the right hand sides 
is being taken in the finite dimensional subspace
of mass level $m$ states only. We shall label the matrices $\wt F_T$ and $\PP_T$ as
\be
\pmatrix{_g\langle r| \wt F_T | r'\rangle_g  & _g\langle r| \wt F_T | s'\rangle_u & 
_g\langle r |\wt F_T | \alpha'\rangle_p\cr
_u\langle s| \wt F_T | r'\rangle_g & _u\langle s| \wt F_T | s'\rangle_u & _u\langle s |\wt F_T | 
\alpha'\rangle_p\cr
~_p\langle\alpha|\wt F_T | r'\rangle_g  &~_p\langle\alpha|\wt F_T | s'\rangle_u 
& ~_p\langle\alpha|\wt F_T | \alpha'\rangle_p
} \quad \hbox{and} \quad
\pmatrix{_g\langle r|  \PP_T | r'\rangle_g  & _g\langle r|  \PP_T | s'\rangle_u & 
_g\langle r | \PP_T | \alpha'\rangle_p\cr
_u\langle s|  \PP_T | r'\rangle_g & _u\langle s|  \PP_T | s'\rangle_u & _u\langle s 
| \PP_T | \alpha'\rangle_p\cr
~_p\langle\alpha| \PP_T | r'\rangle_g  &~_p\langle\alpha| \PP_T | s'\rangle_u & ~_p\langle\alpha| 
\PP_T | \alpha'\rangle_p
}
\ee
respectively.

$\PP_T$ and $\wt F_T$ and the inner product matrix $\II$ are the exact analogs
of the corresponding quantities defined in \S\ref{stoy}. 
In particular the genus expansion of $\wt F_T$ is free from any poles at $k^2=-m^2$
at every order.
In \S\ref{sphprop} we shall
generalize the procedure
of \S\ref{stoy} to construct the propagator of physical states.

One point worth emphasizing is that for our analysis  we do not really need the
gluing compatibility condition discussed in \S\ref{soff} to be valid
for the whole range $0\le s<\infty$
with $s$ defined in \refb{egluing}; it is sufficient if the compatibility condition holds in a small 
neighborhood of degeneration points, {\it e.g.} for $s\ge s_0$ for some constant $s_0$. One way
to see this is that we can rescale all the local coordinates $w_i$
to bring the range $s\ge s_0$  in \refb{egluing}
to $s\ge 0$. This will have the effect of rescaling all the off-shell
amplitudes by some power of
$e^{-s_0(k^2+m^2)}$. But we can also proceed with the original choice of local coordinates and
repeat the whole analysis by changing the definition of 1PI and 1PR amplitudes so that
two or more 
1PI amplitudes glued together using \refb{egluing} for $s\ge s_0$ are now declared as 1PR.
We also have to modify the definition of $\Delta$ given in \refb{eDelta},
with the integral over $s$ now running from $s_0$
to $\infty$. This will produce a multiplicative factor of $e^{-s_0(L_0+\bar L_0)}$ in the
definition of $\Delta$. But the rest of the analysis is not affected by this. In particular we
can continue to define $\bar\Delta$ and $\bar\FF$ via eqs.\refb{edefbardelta} and
\refb{edefbarff}. The contribution to $\bar\Delta$ from states of mass level $m$ now gives
$P_T (k^2+m^2)^{-1} (e^{-s_0(k^2+m^2)/2}-1)$. Since this does not have a pole at
$(k^2+m^2)=0$, $\bar\Delta$ and $\bar\FF$ will continue to be free from poles at
$k^2+m^2=0$.

\subsection{An alternate definition of $\wt F_T$} \label{snewdef}

The definition of $\wt F_T$ given in \S\ref{s4.1} looks complicated, since we first need to define
the 1PI amplitudes $\wh\FF$, then construct $\bar\FF$ via \refb{edefbarff} and finally
project onto the mass level $m$ sector as in \refb{edefppt}. 
In particular the definition of $\wh \FF$ requires dividing up the moduli space of
Riemann surfaces into 1PI and 1PR parts.
Since $\wt F_T$ will play a crucial role in the definition of the physical renormalized mass, 
we shall now give an alternate definition of $\wt F_T$ which does not require 
us to explicitly identify the 1PI subspace in the moduli space of Riemann surfaces.
For this we note from \refb{edefpr} that
\be \label{enewdefwtft}
\wt F_T = F_T (1 + (k^2+m^2)^{-1} \II F_T)^{-1} =
F_T - F_T \II (k^2+m^2)^{-1} F_T + F_T \II (k^2+m^2)^{-1} F_T \II (k^2+m^2)^{-1} F_T +\cdots
\, .
\ee
Now $F_T$ has a simple interpretation since it
denotes the full off-shell 
2-point function restricted to mass level $m$. Thus we can regard
\refb{enewdefwtft} as the definition of $\wt F_T$. In this way of defining $\wt F_T$ we never have to divide
the contribution to an amplitude into 1PI and 1PR parts. The only price we pay is that 
from \refb{enewdefwtft} it is not obvious that $\wt F_T$ is free from poles at $k^2+m^2=0$,
since each term on the right hand side of \refb{enewdefwtft} does contain such poles.
Nevertheless our previous arguments guarantee that all such poles cancel.

It may seem that $\wt F_T$
defined this way requires less information than in the earlier definition, but this is
not the case. The definition of $\wt F_T$ requires information on the choice of local 
coordinate system, which in turn completely fixes the division of the amplitudes into
1PI and 1PR parts. Thus even though we do not explicitly use this division in defining
$\wt F_T$, the data used in the construction of $\wt F_T$ is sufficient to determine
the division of an amplitude into 1PI and 1PR parts.

The definition of $\wt F_T$ given in this subsection will be useful when we generalize the
analysis to super and heterotic string theories.

\subsection{Renormalized physical state propagator and masses} \label{sphprop}

Following the analysis of \S\ref{salgo} we now seek a change of basis
\be \label{ech1string}
|\alpha\rangle'_p = A_{\beta\alpha}|\beta\rangle_p + B_{s\alpha } |s\rangle_g + C_{s\alpha } 
|s\rangle_u, \quad
|r\rangle'_g =  |r\rangle_g + D_{\beta r} |\beta\rangle_p, \quad |r\rangle'_u=|r\rangle_u 
+ K_{\beta r} |\beta\rangle_p\, ,
\ee
such that the following conditions hold
\be \label{econd1string}
'_p\langle \alpha|\beta\rangle'_p =\delta_{\alpha\beta}, \quad 
'_p\langle \alpha|s\rangle'_u=\, '_p\langle \alpha|s\rangle_g'
=\, '_u\langle r|\beta\rangle_p'=\, '_g\langle r|\beta\rangle_p'=0\, ,
\ee
and
\be \label{econd2string}
'_p\langle \alpha|\wt F_T|s\rangle_u'=\, '_p\langle \alpha|\wt F_T|s\rangle_g'
=\, '_u\langle r|\wt F_T|\beta\rangle_p'=\, '_g\langle r|\wt F_T|\beta\rangle_p'=0\, .
\ee
We now substitute \refb{ech1string} into
\refb{econd1string}, \refb{econd2string} and use \refb{eidentitystring} to 
get\footnote{We seek a change of basis that is real in the position space. In momentum
space this implies that changing the momentum from $k$ to $-k$ has the effect of
complex conjugating the coefficients $A_{\alpha\beta},\cdots K_{\beta r}$. This has been
used in \refb{emasterstring}.}
\ben \label{emasterstring}
&& (A^\dagger A + B^\dagger C+C^\dagger B)_{\alpha\beta}
=\delta_{\alpha\beta}, \quad (D^\dagger A + C)_{r\alpha }= 0, \quad (K^\dagger A+B)_{r\alpha }
=0\, , \nonumber \\
&& 
~_u\langle r|\wt F_T|\alpha\rangle_p A_{\alpha\beta} + \, _u\langle r|\wt F_T|s\rangle_g 
B_{s\beta}+\, _u\langle r|\wt F_T|s\rangle_u C_{s\beta} 
+ (K^\dagger)_{r\gamma} 
\, _p\langle \gamma|\wt F_T|\alpha\rangle_p A_{\alpha\beta}  \nonumber \\ &&
\qquad \qquad +  (K^\dagger)_{r\gamma} \, _p\langle \gamma|\wt F_T|s\rangle_g  B_{s\beta}
+ (K^\dagger)_{r\gamma} \, _p\langle \gamma|\wt F_T|s\rangle_u C_{s\beta} 
= 0 \nonumber \\
&& ~_g\langle r|\wt F_T|\alpha\rangle_p A_{\alpha\beta} + \, _g\langle r|\wt F_T|s\rangle_g 
B_{s\beta}+\, _g\langle r|\wt F_T|s\rangle_u C_{s\beta} 
+ (D^\dagger)_{r\gamma} 
\, _p\langle \gamma|\wt F_T|\alpha\rangle_p A_{\alpha\beta}  \nonumber \\ &&
\qquad \qquad +  (D^\dagger)_{r\gamma} \, _p\langle \gamma|\wt F_T|s\rangle_g  B_{s\beta}
+ (D^\dagger)_{r\gamma} \, _p\langle \gamma|\wt F_T|s\rangle_u C_{s\beta} 
= 0 \, .  
\een
Let us first count the number of independent variables and the number of independent
equations. The number of real components in the variables $A_{\alpha\beta}$, 
$B_{s\alpha}$, $C_{s\alpha}$, $D_{\beta r}$ and $K_{\beta r}$ are
\be
2 n_p^2 + 4 \times 2 n_p n_g\, ,
\ee
where $n_p$ is the number of physical states and $n_g=n_u$ is the number of pure gauge
/ unphysical states at mass level $m$. On the other hand the number of independent
equations can be counted as follows. Since both sides of the first equation  in
\refb{emasterstring} are hermitian matrices, this gives $n_p^2$ real equations, whereas
each of the rest gives $2 n_p n_g$ real equations. Thus the total number of equations
is
\be
n_p^2 + 4 \times 2 n_p n_g\, .
\ee
Thus we see that we have $n_p^2$ extra variables compared to the number of equations.
This can be traced to the freedom of multiplying $A$, $B$ and $C$ by a unitary matrix
from the right which is a symmetry of the equations \refb{emasterstring}
(and represent the freedom
of a unitary rotation in the subspace of physical states $|\alpha\rangle'_p$).
Up to this
freedom we can determine the matrices $A,\cdots K$ by solving \refb{emasterstring}.

We shall now describe an iterative procedure for solving these equations. For this we note
that the leading (genus one) contribution to $\wt F_T$ satisfies the property
\be \label{ess.1}
~_p\langle \alpha | \wt F_T| s\rangle_g \sim \lambda\, (k^2+m^2), \quad 
~_g\langle r| \wt F_T| s\rangle_g \sim \lambda\,  (k^2+m^2),  \quad 
~_g\langle r| \wt F_T| \beta\rangle_p \sim \lambda\,  (k^2+m^2)\, ,
\ee
where $\lambda$ now
stands for the genus expansion parameter given by the square of the string coupling.
These properties follow from the fact that at genus one $\wt F_T$ includes the full 
contribution to the torus two point function. Representing a pure gauge state as $Q_B|n\rangle$
plus a term of order $(k^2+m^2)$, deforming the contour
of integration of the BRST current so that it acts on the
other vertex operator, and then using that fact that acting on an off-shell physical or pure
gauge state $Q_B$ gives a term proportional to $(k^2+m^2)$, we arrive at \refb{ess.1}.
This in turn allows us to look for solutions where at order $\lambda^0$,
\be \label{eorderstring}
A, B, K \sim 1,  \quad C,D\sim (k^2+m^2) \, .
\ee
The solution to order $\lambda^0$ and leading order in $k^2+m^2$ are given by
\ben \label{esolution}
&& A_{\alpha\beta} =\delta_{\alpha\beta} +\OO(k^2+m^2), \quad C = - D^\dagger+\OO(k^2+m^2), 
\quad B = -K^\dagger +\OO(k^2+m^2), 
\nonumber \\
&& \lambda^{-1}\, \{ \delta_{\beta\gamma}\, _u\langle r|\wt F_T|s\rangle_g  
-  \delta_{rs}\, _p\langle \gamma|\wt F_T|\beta\rangle_p\}  K^\dagger_{s\gamma} 
= \lambda^{-1}\,  \, _u\langle r|\wt F_T|\beta\rangle_p +\OO(k^2+m^2)\nonumber \\ &&
\lambda^{-1}\,  \{ \delta_{\beta\gamma}\, _g\langle r|\wt F_T|s\rangle_u
-  \delta_{rs}\, _p\langle \gamma|\wt F_T|\beta\rangle_p\}  D^\dagger_{s\gamma} 
\nonumber \\
&& \qquad =\lambda^{-1}\, \, _g\langle r|\wt F_T|\beta\rangle_p  + \lambda^{-1}\,  
\, _g\langle r|\wt F_T|s\rangle_g \, B_{s\beta}
+\OO\left((k^2+m^2)^2\right)\, . 
\een 
This gives a sensible solution satisfying \refb{eorderstring} 
provided the $n_pn_g\times n_pn_g$ matrix
\be \label{emainmatrix}
S_{r\beta, s\gamma} \equiv \lambda^{-1} \{ \,
_u\langle r|\wt F_T|s\rangle_g \,  \, \delta_{\beta\gamma} 
-  \, _p\langle \gamma|\wt F_T|\beta\rangle_p \, \, \delta_{rs}\},
\ee
is invertible. Starting with this solution we can solve for the matrices $A,B,C,D,K$
iteratively in powers of the genus expansion parameter $\lambda$ and $(k^2+m^2)$
exactly as in \S\ref{stoy}. As
long as the matrix defined in \refb{emainmatrix} is invertible, the coefficient of $\lambda^n$
for any $n$ is free from poles near $k^2=-m^2$. Physically, invertibility of 
$S_{r\beta,s\gamma}$ is the condition that the degeneracy between the masses of physical
states and the unphysical / pure gauge states is lifted at one loop order. If this condition fails
then we need to go to higher order in perturbation theory to lift the degeneracy. We expect that
in principle there should be no difficulty in carrying out this procedure, although in practice the
analysis is likely to become more complicated.

The coefficients $A,\cdots K$ satisfying \refb{emasterstring} ensures, via 
eqs.\refb{ech1string}-\refb{econd2string} that the matrices $\II$ and $\wt F_T$ expressed in the
primed basis have block diagonal form, with no cross terms between the states
$|\alpha\rangle'_p$ and ($|r\rangle'_u$, $|r\rangle'_g$). As in \S\ref{stoy} we define
\be \label{edefwtf}
\wt F_{\alpha\beta}(k) = \, '_p\langle \alpha| \wt F_T|\beta\rangle'_p\, .
\ee
Then the propagator restricted to the modified physical sector is given by
\be \label{eprophys}
\PP_{\alpha\beta}  \equiv  \, '_p\langle \alpha| \PP_T|\beta\rangle'_p
= \left((k^2 + m^2 - \wt F(k))^{-1}\right)_{\alpha\beta}\, .
\ee
From here onwards we proceed as in \cite{1311.1257}.
We can diagonalize $\wt F(k)$ as
\be 
\wt F(k) = U(k) \wt F_d(k) U(k)^\dagger, \qquad U(k)^\dagger = U(k)^{-1} = U(-k)^T\, ,
\ee
so that we have
\be \label{epp}
\PP = U(k) (k^2 + m^2 - \wt F_d(k))^{-1} U(k)^\dagger\, .
\ee
We can now determine the solutions to the equation $k^2+m^2- \wt F_d(k)=0$ iteratively
for each of the diagonal entries of $\wt F_d(k)$, starting with $k^2=-m^2$ as the leading
order solution. This gives the physical masses. Let $M_p^2$ denotes the diagonal
matrix with the diagonal elements being equal to the squares of the
physical masses. Then we can
express $(k^2+m^2 -\wt F_d(k))^{-1}$ as
\be 
X_d(k) (k^2+M_p^2)^{-1}\, ,
\ee
where $X_d(k)$ is a diagonal matrix which has no poles near $k^2=-m^2$. 
Eq.\refb{epp} now allows us to
express the physical propagator $\PP_{\alpha\beta}$ as
\be
\PP = Z^{1/2}(k) (k^2+M_p^2)^{-1} Z^{1/2}(-k)^T, \qquad
Z^{1/2}(k) \equiv U(k) X_d(k)^{1/2}\, .
\ee
In \S\ref{sall} we shall argue that the squares of the
physical masses given by the diagonal
elements of $M_p^2$
do not depend on the choice of local coordinates at the punctures, 
although the wave-function renormalization
matrix $Z^{1/2}(k)$ does depend on the choice of local coordinates.

Finally we would like to note that since 
$B_{s\alpha}$ is of order unity, the corrected physical state
$|\alpha\rangle'_p$ differs from the tree level physical state
$|\alpha\rangle_p$ by a pure gauge state with coefficient of order
unity. Thus even in the $\lambda\to 0$ limit, $|\alpha\rangle_p'$ does 
not approach $|\alpha\rangle_p$. 

\subsection{Renormalized masses in the unphysical / pure gauge sector} \label{eunpstring}

We shall now briefly describe the computation of the renormalized masses in the
unphysical / pure gauge sector by generalizing the procedure described in \S\ref{sunp2}.
For this we define
\be \label{eupup}
\II' = \pmatrix{~_g'\langle r| s\rangle'_g & ~'_g\langle r| s\rangle'_u\cr
~'_u\langle r| s\rangle'_g & ~'_u\langle r| s\rangle'_u}, \quad
\wt F' = \pmatrix{~_g'\langle r| \wt F_T|s\rangle'_g & ~'_g\langle r|  \wt F_T| s\rangle'_u\cr
~'_u\langle r|  \wt F_T| s\rangle'_g & ~'_u\langle r|  \wt F_T| s\rangle'_u}\, .
\ee
Then the renormalized mass$^2$'s in the unphysical / pure gauge sector will be given by the
zeroes of the eigenvalues of the matrix
\be \label{eunmatrix}
(k^2 + m^2) \II' - \wt F'(k)\, ,
\ee
in the complex $-k^2$ plane. To evaluate the order $\lambda$ correction to these masses, we
shall assume as in \S\ref{sunp2} that $k^2+m^2$ is of order $\lambda$ when $-k^2$ is equal
to the renormalized 
mass$^2$
and keep terms in \refb{eunmatrix} up to order $\lambda$. Using \refb{ech1string},
\refb{ess.1} and \refb{esolution} one finds that to order unity
\be  ~_g'\langle r| s\rangle'_g  = 0, \quad  ~_g'\langle r| s\rangle'_u  = 
~_u'\langle r| s\rangle'_g  = \delta_{rs}, 
\ee
and to order  $\lambda$,
\be ~_g'\langle r| \wt F_T|s\rangle'_g = 0,
\quad ~_g'\langle r| \wt F_T|s\rangle'_u = ~_g\langle r| \wt F_T|s\rangle_u,
\quad  ~_u'\langle r| \wt F_T|s\rangle'_g = ~_u\langle r| \wt F_T|s\rangle_g \, .
\ee
Hence to order $\lambda$ (counting $k^2+m^2$ as order $\lambda$)
\be
(k^2 + m^2) \II' - \wt F'(k) =
\pmatrix{ 0 &  (k^2 + m^2)  \, \delta_{rs} - ~_g\langle r|  \wt F_T| s\rangle_u\cr
 (k^2 + m^2) \,  \delta_{rs}  - ~_u\langle r|  \wt F_T| s\rangle_g& 
  (k^2 + m^2)  ~'_u\langle r| s\rangle'_u
- ~'_u\langle r|  \wt F_T| s\rangle'_u} \, .
\ee
Using the fact that the vanishing of an eigenvalue of a matrix is equivalent to requiring 
the vanishing of its determinant, we see that the required condition is the vanishing of the
determinant of the upper right (or lower left) block. This in turn is equivalent to requiring the
vanishing of an eigenvalue of
\be  \label{efinunp}
(k^2 + m^2)  \,  \delta_{rs} - ~_g\langle r|  \wt F_T| s\rangle_u
\ee
as a function of $-k^2$. Starting with this first order solution one can iteratively
compute higher order corrections to the renormalized mass$^2$ in the unphysical /
pure gauge sector by looking for zero eigenvalues of \refb{eunmatrix}.

\subsection{Dependence on choice of local coordinates} \label{sdepend}

An important question is: how do the physical masses depend on the choice of
local coordinates? We shall postpone a full discussion on this till \S\ref{sall}, but at this stage
we can derive the result at order $\lambda$. 
The locations of the physical mass squares are determined by the zeroes of  %as
$k^2+m^2 - \wt F_d(k)$
in the $-k^2$ plane.
Let us focus on the one loop, \i.e.\ order $\lambda$ 
correction to the mass$^2$.
For this we need to determine the function $\wt F_d(k)$ and hence 
$\wt F(k)$ to order $\lambda$ at
$k^2=-m^2$.  It follows from \refb{ech1string}, \refb{ess.1}, \refb{esolution},
\refb{edefwtf} and the fact that the leading
contribution to $\wt F_T$ is of order $\lambda$ that to order $\lambda$ and at $k^2+m^2=0$
\be
\wt F_{\alpha\beta} = \langle \alpha| \wt F_T | \beta \rangle \, .
\ee
At order $\lambda$ this represents the full two point function of the tree level physical states
$|\alpha\rangle$ and $|\beta\rangle$ on the torus.  Since $|\alpha\rangle$ and $|\beta\rangle$
are both dimension zero primaries at $k^2=-m^2$, we see that to this order
$\wt F_{\alpha\beta}$ at $k^2=-m^2$ is independent of the choice of local coordinates. Hence
the renormalized physical masses are also independent of the choice of local coordinates to this
order.

We can also consider the fate of the masses in the unphysical / pure gauge sector under a
change in the local coordinate system. 
To order $\lambda$ the  mass$^2$'s in this sector are given by the zeroes of the eigenvalues of
the matrix \refb{efinunp} in the $-k^2$ plane.
Since the matrix elements $ ~_g\langle r|  \wt F_T| s\rangle_u$ 
involve unphysical and pure
gauge states, which are generically not dimension zero primaries,
we see that  in the generic case the
order $\lambda$ contribution to the masses of the unphysical and pure gauge states will
depend on the choice of local coordinates.\footnote{If the vertex operator
involves ghost excitations then the integration measure provided by $b$-ghost insertions also
depend on the choice of local coordinates\cite{Nelson,9206084}.} Higher order 
contributions can correct these
results but cannot cancel the order $\lambda$ corrections. This we conclude that the unphysical
/ pure gauge sector masses do depend on the choice of local coordinate system.

\sectiono{Poles of S-matrix elements of massless / BPS / special states} \label{spoles}

In this section 
we shall show that
if we consider an S-matrix of external massless, BPS and/or special states then the poles
in this S-matrix in any channel are the same ones as those which appear in the analysis
of \S\ref{sphprop}.\footnote{This generalizes the result of 
\cite{Polchinski:1988jq} in the absence of mass
renormalization.}
Let us denote  by $k$ the total momentum carried in some particular internal channel, being
equal to the sum of momenta of two or more external states, and look for poles in the $-k^2$
plane.
Our starting point will be the expression \refb{eGamma} for the $(m+n)$-point amplitude.
The S-matrix elements are obtained from this by multiplying this by appropriate 
renormalization factors on the
external legs and then setting the external momenta on-shell. Since multiplicative factors
on the external legs do not affect
the locations of the poles in the $k^2$ plane, we can directly use $\Gamma$ to examine these poles.
Our interest will be to look for those poles which arise from states at mass level
$m$. For this it will be useful to `integrate out' the states at other mass levels. With this
goal in mind, we define
\ben\label{edefbargamma12}
\bar\Gamma_1^T\II &=& \wh\Gamma_1^T \II (1 + \bar\Delta \wh\FF + \bar\Delta \wh\FF \bar
\Delta \wh\FF +\cdots)
= \wh\Gamma_1^T \II (1 - \bar\Delta \wh\FF)^{-1}\, , \nonumber \\
\bar\Gamma_2 &=& (1 +  \wh\FF\bar\Delta+ \wh\FF\bar\Delta \wh\FF\bar\Delta 
+\cdots) \, \wh\Gamma_2 = (1 -  \wh\FF\bar\Delta)^{-1} \, \wh\Gamma_2\, ,
\een
where $\bar\Delta$ has been defined in \refb{edefbardelta}.
We also define
\be\label{edefbargamma}
\bar\Gamma = \wh\Gamma + \wh\Gamma_1^T \II (\bar\Delta + \bar\Delta \wh\FF \bar\Delta
+  \bar\Delta \wh\FF \bar\Delta\wh\FF \bar\Delta + \cdots) \wh\Gamma_2
= \wh\Gamma + \wh\Gamma_1^T \II \bar\Delta (1 - \wh\FF \bar\Delta)^{-1} \wh\Gamma_2
= \wh\Gamma + \wh\Gamma_1^T \II  (1 -\bar\Delta \wh\FF)^{-1} \bar\Delta \wh\Gamma_2\, .
\ee
Using \refb{eGamma}, \refb{edefbardelta}, \refb{edefbarff}, \refb{edefbargamma12} 
and \refb{edefbargamma} we now get
\ben \label{e4.3}
\Gamma &=& \bar\Gamma + \bar\Gamma_1^T\II \left\{1 - (k^2+m^2)^{-1} P_T \bar\FF\right\}^{-1} 
P_T (k^2+m^2)^{-1}
\bar\Gamma_2  \nonumber \\
&=& \bar\Gamma + \bar\Gamma_1^T\II P_T (k^2+m^2)^{-1} \left\{1 - (k^2+m^2)^{-1}
\bar \FF P_T\right\}^{-1} \bar\Gamma_2\nonumber \\
&=& \bar\Gamma + \bar\Gamma_1^T\II  P_T (k^2 + m^2 - P_T \bar\FF P_T)^{-1} P_T \bar\Gamma_2
\nonumber \\
&=& \bar\Gamma + \bar\Gamma_1^T \PP_T \bar\Gamma_2 \, ,
\een
where $\PP_T$ has been defined in \refb{edefppt}.
Now the genus expansions of 
$\bar\Gamma$, $\bar\Gamma_1^T$ and $\bar\Gamma_2$ are free from poles at
$-k^2=m^2$. Thus the only poles near $-k^2=m^2$ can come from the poles of
matrix $\PP_T$. These are precisely the renormalized
physical and unphysical squared masses as discussed
in \S\ref{sphysical}. 

For later use, it will be useful 
to isolate the contribution from the physical states from that of the unphysical and
pure gauge states. For this we insert the projection
operator $P_T$ on both sides of $\PP_T$ on the right hand side of
\refb{e4.3} using the identity $P_T \PP_T P_T =\PP_T$. 
Now using \refb{edefpt} and \refb{ech1string}, $P_T$ may be
expressed as
\be \label{exx23}
P_T =  \sum_\alpha |\alpha\rangle'_p  ~_p'\langle \alpha| +
\sum_{r,s} \left[\wt A_{rs} |r\rangle_g' ~_g'\langle s| + \wt B_{rs} |r\rangle_g' ~_u'\langle s| +
\wt C_{rs} |r\rangle_u' ~_g'\langle s| + \wt D_{rs} |r\rangle_u' ~_u'\langle s| \right]\, ,
\ee
where $\wt A_{rs}$, $\wt B_{rs}$, $\wt C_{rs}$ and $\wt D_{rs}$ are constants which can be
computed from \refb{ech1string}, \refb{emasterstring}. The first term on the right hand side of
\refb{exx23} describes the contribution from renormalized 
physical states whereas the other terms represent
contribution from renormalized unphysical and pure gauge states.

Let us now examine the residues at the poles in \refb{e4.3}
at leading order in string perturbation theory.
First consider the residue at a physical pole. This
is given by the
products of 
the components of $\bar\Gamma_1$ and $\bar\Gamma_2$ along the corresponding physical 
state $|\alpha\rangle'_p$.
At
the tree level the relevant component of 
$\bar\Gamma_1$ is given by the contribution to the full $(m+1)$ point tree 
amplitude  with external states $|\alpha\rangle'_p$ and $m$
other massless / BPS / special states, and
similarly the relevant component of 
$\bar\Gamma_2$  is given by the contribution to the full $(n+1)$ point tree 
amplitude  with external states $|\alpha\rangle'_p$ and $n$ 
other massless / BPS / special states.
Since in the leading order $|\alpha\rangle'_p$ is given by
a linear combination of tree level physical state $|\alpha\rangle_p$
and a pure gauge state, and since the 
pure gauge states decouple in the on-shell tree level amplitude, we can replace
$|\alpha\rangle'_p$ by $|\alpha\rangle_p$ in computing the leading order
contribution to the relevant components of 
$\bar\Gamma_1$ and $\bar\Gamma_2$. 
Thus in the leading order the residue at the physical pole is given by the
product of  two tree level S-matrix elements -- one with $(m+1)$ external states and the
other one with $(n+1)$ external states. As long as these are non-zero, the residue at the
corresponding physical pole will be non-zero. Higher order contributions can correct the
residue but cannot make this vanish in perturbation theory. Thus  even after
including higher order corrections, the corresponding
physical mass$^2$'s will appear as the locations of the poles in the $-k^2$ plane
of the original S-matrix element involving $(m+n)$ external massless / BPS / special states.

Let us now turn to the contribution from the unphysical / pure gauge states. It follows from
\refb{ech1string}, \refb{ess.1}, \refb{esolution} and \refb{exx23} that for $k^2=-m^2$ 
and leading order in $\lambda$, 
the coefficients  $\wt D_{rs}$ vanish. On the other hand the same equations show that in
this approximation $|s\rangle_g'=|s\rangle_g$.
Thus the residue is given by a sum of products of appropriate components of
$\bar\Gamma_1$ and $\bar\Gamma_2$, and in each of these terms either the component of
$\bar\Gamma_1$ or the component of $\bar\Gamma_2$ (or both) is aligned along
a tree level pure gauge state $|s\rangle_g$.
Thus this factor is given by a tree level amplitude, one of whose external states is
$|s\rangle_g$ and the other states are on-shell massless / pure gauge / special states.
Since this vanishes due to
BRST invariance, we conclude
that at least at leading order in $\lambda$ the unphysical states do not contribute to the poles
in the S-matrix elements of massless / BPS / special states.

Before concluding this section we would like to note that the various quantities which appear in 
\refb{e4.3} -- {\it e.g.} $\bar\Gamma$, $\bar\Gamma_1^T \II P_T$, $P_T \bar\Gamma_2$
etc. -- can be defined without having to explicitly identify the 1PI Riemann surfaces by following
the same strategy as in \S\ref{snewdef}. 
For example we have the relations
\ben \label{enewrefgamma}
\bar\Gamma_1^T \II P_T &=& \Gamma_1^T\, \II \, (1 - (k^2+m^2)^{-1}  \II \wt F_T) \, P_T \, ,
\nonumber \\
P_T \bar \Gamma_2 &=&  (1 -  (k^2+m^2)^{-1}   \II\wt F_T) \, P_T\, \Gamma_2\, , \nonumber \\
\bar\Gamma &=& \Gamma - \Gamma_1^T \, \II\,  (k^2+m^2)^{-1}\,
(1 -  (k^2+m^2)^{-1}   \II \wt F_T) \, P_T\, \Gamma_2\, .
\een
Since $\Gamma$, $\Gamma_1$ and $\Gamma_2$ are full amplitudes, their
definitions do not require us to divide the moduli space of Riemann surfaces into 1PI 
and 1PR parts. The definition of $\wt F_T$ given in \S\ref{snewdef} also does not
require this division. Thus $\bar\Gamma_1^T \II P_T$, $P_T\bar\Gamma_2$ and $\bar \Gamma$
defined via \refb{enewrefgamma} also do not require this divison.
This observation will be useful when we generalize the
analysis to super and heterotic string theories.

\sectiono{All order results} \label{sall}

We shall now combine the results of \S\ref{sdepend} and \S\ref{spoles}  to prove
some all order results in a generic situation. For this we need to first explain what we mean by
a generic situation. The conditions under which our arguments will hold are listed below.
\begin{enumerate}
\item We assume that the degeneracies %as 
between physical and unphysical masses are lifted at first
order in perturbation theory. Otherwise our prescription of \S\ref{sphysical} of computing renormalized
physical masses will have to be modified.
\item We have seen that to leading order the residue at a particular physical mass$^2$ of an S-matrix
element of external massless / BPS / special states is proportional to the product of S-matrix
elements of two lower point tree level S-matrix elements each of which contains, as 
one of the external states,
the physical 
state whose mass
we are interested in. We shall assume that it is possible to choose the
external massless / BPS / special states of the original amplitude in such a way that 
both these lower point S-matrix elements are non-vanishing at tree level.
Had we restricted the external
states to be only massless or BPS states then this fails in some cases, as was illustrated in 
\cite{1311.1257}. (A particular example of this is the SO(32) spinor states of ten dimensional
SO(32) heterotic
string theory; these cannot appear as one particle intermediate states in the scattering of massless
external states which are all in the adjoint or singlet representation of SO(32).)
However at present we do not know of an example where it fails even after
we allow as external states the special states introduced in \cite{1311.1257}.
Once the residue at the pole can be made non-vanishing at leading order, higher order corrections
can modify the residue but cannot make it vanish in perturbation theory.
\item We have seen in \S\ref{sdepend} that the renormalized masses of
unphysical / pure gauge states do in general depend on the choice of local coordinates. We shall
assume that this is true in all cases, \i.e.\ there is no renormalized mass corresponding to
unphysical / pure gauge states which is accidentally independent of the choice of 
local coordinates.
\end{enumerate}

Next we shall combine the genericity assumption with some of the relevant results in 
\S\ref{sdepend},
\S\ref{spoles} and ref.\cite{1311.1257} to draw the following conclusions: 
\begin{enumerate}
\item In a generic situation the renormalized 
masses of the unphysical / pure gauge states   depend on the choice of local coordinates.
\label{point1}
\item It was shown in \S\ref{sdepend} the renormalized masses of physical states
do not
depend on the choice of local coordinates at least to order $\lambda$. \label{point2}
\item In a generic situation the mass$^2$ of physical states 
appear as poles in the $-k^2$ plane of some
S-matrix of massless / BPS / special states. \label{point3}
\item It was also shown near the end of \S\ref{spoles} that 
the unphysical /pure gauge states do not contribute poles in the 
S-matrix of massless / BPS / special states
at least to leading order in $\lambda$. \label{point4}
\item The S-matrix involving external massless / BPS / special states do not
depend on the choice of local coordinates to all orders in $\lambda$\cite{1311.1257}. 
\label{point5}
\end{enumerate}

Let us now combine these results. 
Points \ref{point1} and \ref{point5} show, to all orders in $\lambda$, 
that the unphysical / pure gauge states cannot appear as
intermediate states in the S-matrix of massless / BPS / special states.
This is consistent with the leading order result mentioned in point \ref{point4}.
On the other hand points \ref{point3} and \ref{point5} show, to all orders in
$\lambda$, that the mass$^2$ of
physical states cannot depend on the choice of local coordinates. This is
consistent with the leading order result described in point \ref{point2}.
 
We can also extend this argument to prove the invariance of the S-matrix elements
of general external physical states under a change of local coordinates. For this we note
that as long as each of the external states have non-zero tree level amplitude with some
set of massless / BPS / special states, we can replace each of the massive,
non-BPS and non-special external 
physical states by the corresponding combination of massless / BPS / special states and
examine the corresponding S-matrix for values of momenta where the intermediate
physical states of interest go on-shell. The desired 
S-matrix can then be found by examining the 
residue at the pole.\footnote{If the physical state under consideration is unstable 
then this is the only way to
define its `S-matrix' since the state does not exist as asymptotic state.} Since the
S-matrix of massless / BPS / special states is invariant under a change in the local coordinate
system, its residues at various poles must also be invariant under a change of local coordinates.
This establishes the invariance of the S-matrix elements involving general external physical
states under a change of the local coordinate system.

\sectiono{Generalizations to heterotic and super string theories} \label{shetsup}

We shall now briefly discuss generalizations to heterotic and superstring theories.
We shall restrict our discussion to the Neveu-Schwarz (NS) sector, and 
work with picture number $-1$ states. 
In this case the discussion of \S\ref{sorganize} can be adapted with few changes:
\begin{enumerate} 
\item The discussion in \S\ref{sbasis} remains valid without any change. 
\item In the
analysis of \S\ref{soff} we need to choose local superconformal coordinate system 
$(w,\xi)$ around every puncture for defining off-shell amplitudes, and $f_i\circ V_i$ will
label the transform of the vertex operator $V_i$ by the superconformal transformation
$f_i$ that
relates the local coordinates near the $i$-th puncture to the reference superconformal
coordinates on the super Riemann surface. The detailed analysis of the integration measure
(ghost insertions) for off-shell amplitudes
can be carried out by combining the
description of the measure for on-shell amplitudes in super and heterotic string theories
given in \cite{1209.5461}  with the description of the measure for off-shell amplitudes in bosonic
string theory given in \cite{Nelson,9206084}.
\item
The gluing of two
Riemann surfaces is implemented via the identification\cite{1209.5461}
\be
w_1 w_2 = q_{NS}, \quad w_2 \xi_1 =\ve\xi_2,
\quad w_1\xi_2 = -\ve\xi_1, \quad \xi_1\xi_2=0, \quad
\ve = \pm \sqrt{-q_{NS}}\, ,
\ee
and we need to sum over both choices of the sign of $\ve$, leading to GSO
projection.
\item The choice of local superconformal coordinates should be compatible with
gluing in the same way as in the case of bosonic string theory. 
We shall also require that the contours in the supermoduli space over which we 
integrate\cite{1209.5461}
should be compatible with gluing.
In particular this means that
in situations where we can integrate out the odd moduli at the expense of inserting picture
changing operators\cite{FMS,VERLINDE}, the locations of the picture changing operators on the glued
Riemann surface should be those induced from the locations of the picture changing
operators on the lower genus Riemann surfaces which are being glued. A consistent super or
heterotic string field theory should automatically satisfy this 
property in a Siegel like gauge. Construction of classical superstring field theory satisfying these
requirements can be found in \cite{9202087,0409018}.
\item We can now define $\wt F_T$ following the procedure outlined in \S\ref{snewdef}. This
avoids having to divide the super Riemann surfaces into 1PI and 1PR surfaces, and directly gives
us the expression for $\wt F_T$ in terms of the full off-shell two point function $F_T$ of mass level
$m$ states. Similarly generalization of the analysis of \S\ref{spoles} can also be carried out by
defining $\bar\Gamma$, $\bar\Gamma_1^T \II P_T$, $P_T \bar\Gamma_2$
etc. as in \refb{enewrefgamma} instead of in terms of 1PI super Riemann
surfaces.
\end{enumerate}
The rest of the analysis can be carried out in a straightforward matter
and we arrive at the same conclusions as in the case
of bosonic string theory.

The difficulty in the Ramond sector stems from the fact that there is no natural inner product
between states in the $-1/2$ picture since the inner product pairs states in the $-1/2$ picture
to states in the $-3/2$ picture. Thus generalization of the analysis of \S\ref{sbasis} will
require us to work with picture number $-1/2$ and $-3/2$ states together. On the other hand
superstring perturbation theory naturally uses $-1/2$ picture vertex operators\cite{1209.5461}.
This seems to be a technical issue which needs to be addressed, possibly by introducing a
$\delta(\gamma_0)$ in the definition of the inner product \refb{einnerp} so as to get a 
non-vanishing
inner product between two picture number $-1/2$ states.  We hope to return to
this issue in the future.

\bigskip

{\bf Acknowledgement:}
We thank Rajesh Gopakumar, Michael Green and Edward Witten  for useful discussions, 
and Barton Zwiebach for useful discussions and critical comments on an earlier version
of the manuscript.
We would 
like to 
thank IIT, Kharagpur and the National Strings Meeting held there for hospitality during
the course of of this work.
A.R. would   like to
thank HRI, Allahabad for hospitality during the initial and final stages of this work.
A.S. would like to thank the organizers of the National Strings Meeting in IIT Kharagpur,
MathPhys 2014 conference at Rykkyo University and Asian Winter School 
at Puri, where preliminay version of this work was presented.
The work of R.P. and A.S. was
supported in part by the 
DAE project 12-R\&D-HRI-5.02-0303. 
A.R. was supported by the Ramanujan studentship of Trinity College, 
Cambridge, and his research leading to these results 
has also 
received funding from the European Research Council under 
the European Community's Seventh Framework Programme 
(FP7/2007-2013) / ERC grant agreement no. [247252].
The work of A.S. was also supported in
part by the
J. C. Bose fellowship of 
the Department of Science and Technology, India.

%\small \baselineskip 12pt

\end{document}